\documentclass[letterpaper]{JHEP3}
\usepackage{array}
\usepackage{nicefrac}
\usepackage{amsmath}
\usepackage{amssymb,amsfonts,mathrsfs}
\usepackage{graphicx}
\usepackage{epsfig}

\usepackage{amsfonts}

\def\t{\tau}
\def\a{\alpha}

\def\h{\eta}

\def\half{{\frac12}}

\def\IC{\relax\hbox{$\inbar\kern-.3em{\rm C}$}}

\def\IC{{\bf C}}

\def\bea{\begin{eqnarray}}
\def\eea{\end{eqnarray}}
\def\be{\begin{equation}}
\def\ee{\end{equation}}
\def\ba{\begin{align}}
\def\ea{\end{align}}
\def\bse{\begin{subequations}}
\def\ese{\end{subequations}}
\def\1F1{{}_1\!F_1}
\def\2F0{{}_2\!F_0}

\def\a{\alpha}

\def\h3{$\textrm{H}_3^+$}

\def\IC{{\mathbb C}}

\def\lbldef#1#2{\expandafter\gdef\csname #1\endcsname {#2}}

\def\href#1#2{#2}

\newcommand{\beq}{\begin{equation}}
\newcommand{\eeq}{\end{equation}}
\newcommand{\ber}{\begin{eqnarray}}
\newcommand{\eer}{\end{eqnarray}}

\def\be{\begin{eqnarray}}
\def\ee{\end{eqnarray}}

\def\({\left(}
\def\){\right)}
\def\[{\left[}
\def\]{\right]}
\def\<{\langle}
\def\>{\rangle}

\preprint{NSF-KITP-12-040}

\title{Exceptional Indices}
\author{Davide Gaiotto$^{\diamondsuit}$
 and Shlomo S. Razamat$^{\diamondsuit,\clubsuit}$
\\
\\
$^{\diamondsuit}$ \it Institute for Advanced Study, Princeton, NJ 08540, USA\\
$^{\clubsuit}$ \it Kavli Institute for Theoretical Physics, Santa Barbara, CA 93106, USA\\
}

\bigskip

\abstract{
Recently a prescription to compute the superconformal index for 
\textit{all} theories of class ${\cal S}$ was proposed. In this paper we 
discuss some of the physical information which can be extracted from
this index. We derive a simple criterion for the given theory 
of class ${\cal S}$ to have a decoupled free component and for it to have 
enhanced flavor symmetry.  Furthermore, we establish a criterion for the 
``good'', the ``bad'', and the ``ugly'' trichotomy of the theories. After interpreting the 
prescription to compute the index with non-maximal flavor symmetry as 
a residue calculus we address the computation of the index of the bad theories.
In particular we suggest explicit expressions for the superconformal index
of higher rank theories with $E_n$ flavor symmetry, i.e. for the Hilbert series 
of the multi-instanton moduli space of $E_n$.   
}

\begin{document}

\section{Introduction}

A very rich and tractable arena for obtaining exact results for
quantum field theories in four dimensions is given by
the ${\cal N}=2$ supersymmetric conformal field theories. In recent years
a certain set of these theories, obtainable by compactifications of $M5$ branes
on Riemann surfaces (called theories of class ${\cal S}$~\cite{Gaiotto:2009we, Gaiotto:2009hg}), has received special attention.
  The six dimensional origin  equips these theories
with several very interesting  properties such as generalized S-duality~\cite{Gaiotto:2009we}
and the AGT relation~\cite{Alday:2009aq}.

An example of an analytic computation which can be performed for theories of 
class ${\cal S}$ is the evaluation of the superconformal index~\cite{Kinney:2005ej,Romelsberger:2005eg}.
The superconformal index for vast majority of these theories is completely
fixed by merely assuming  the existence of  S-duality interconnecting them~\cite{Gadde:2011uv,GRR}.  
The purpose of this paper is to extend the prescription for computing the index
to an additional sub-class of theories of class ${\cal S}$ not 
connected to others by S-duality transformations.  
Some of these theories have a Lagrangian description and thus the computation 
of the index is straightforward.  In other cases these theories will be strongly-coupled
SCFTs with no direct way to compute the index.
A simplest example of the former is the ${\cal N}=4$ SYM and an example of the latter
is rank 2 SCFT with $E_6$ flavor symmetry.  

The prescription of~\cite{Gadde:2011uv} to compute the superconformal index of theories of class ${\cal S}$ can be shown~\cite{GRR} 
to follow directly from S-duality properties of the index of the
underlying theories~\cite{Gadde:2009kb}. This prescription translates the data needed to define the compactification
of the $(2,0)$ theory on a Riemann surface to the index. However, for some compactifications it gives
divergent results. These compactifications, defined by a  Riemann surface with punctures, have a common feature that they can not 
be glued to other theories, and thus S-duality can not be used to fix their index. On the other hand, for some of these 
compactifications it is widely believed that a well behaved four dimensional theory exists. The basic example again is
a torus without punctures which is believed to describe ${\cal N}=4$ SYM. 

The divergence of the index calculation can be ascribed to the presence of operators with an unexpected assignment of R-charge. 
In turns, this signals a breakdown of the assumption that the UV R-charge assignment persists in the IR. 
Similar phenomena were encountered in three-dimensional ${\cal N}=4$ SCFTs~\cite{Gaiotto:2008ak},
concerning the R-charge assignment of BPS monopole operators. In \cite{Gaiotto:2011xs} it was argued that 
much the same problem could affect four-dimensional theories in class ${\cal S}$, concerning the BPS operators 
which parameterize the Higgs branch.  

We will call  the theories for which the index diverges ``bad''. 
In this paper we start by giving a simple and precise criterion when the index computation fails to converge.
We further define theories as ``ugly''
if they contain decoupled free matter: again we give a simple criterion for this to happen. Finally, theories which
are not bad nor ugly are called ``good'' ones. 

Further, we reinterpret the prescription to compute the index as an iterative process. Starting with the index of a theory
with all maximal punctures, index of theories with reduced flavor symmetry is obtained by computing residues of the former.
This type of residue calculus was introduced in~\cite{GRR} and is based on the intuition that 
the theories with non-maximal punctures live ``at infinity'' in the Higgs branch of the theories 
with maximal punctures. The specific direction at infinity is selected by looking at the action of the flavor symmetry. 

We observe next that in this procedure, at least in some cases, the bad theories are obtained from ugly ones.
The index of the ugly theories is  well defined and physically sensible. The contribution to the index of the decoupled 
free hypermultiplet turns out to be directly responsible for the singularity which appears at the next step of the procedure. 
We can sharpen our prescription if we distinguish the flavor symmetry rotating the free hypermultiplets from the flavor symmetry rotating 
the interacting part of the SCFT. This will allow us to compute a well-defined index for some of the bad theories. 

In particular we suggest explicit expressions for the index of higher rank theories with $E_n$ flavor symmetry.
These theories are bad in our classification but can be obtained from ugly ones. The higher rank $E_n$ theories
are of particular interest since their Higgs branch is believed to coincide with the moduli space of multi-instantons 
of $E_n$ (see e.g.~\cite{Benini:2009gi,Moore:2011ee}). This moduli space is not very well studied since the ADHM construction here is
lacking. A version of the index on which we will concentrate in this paper, the Hall-Littlewood index in the notations of~\cite{Gadde:2011uv}, is equal to the Hilbert series
of the Higgs branch for quivers associated to genus zero Riemann surfaces. Thus,
we conjecture that  the index we compute counts holomorphic 
functions on the moduli space of multi-instantons. Analogous quantity for a single instanton was computed in~\cite{Garfinkle,Benvenuti:2010pq}
and matches the index computation of~\cite{Gadde:2010te,Gadde:2011uv}.
 
\

This paper is organized as follows. In section~\ref{boxsec} we  review the prescription
of~\cite{Gadde:2011uv} to compute the index of theories of class ${\cal S}$. We define the trichotomy
between the good, the bad, and  the ugly; and then reformulate the prescription as 
a residue computation. Then in section~\ref{uglysec} we discuss theories for 
which special care has to be taken in applying the prescription.  In particular we give an explicit expression for the 
index of the rank two SCFT with $E_6$ flavor symmetry. An appendix contains additional technical details 
of results and claims presented in the bulk of the paper.

\section{Reducing the flavor symmetry ``one box at a time''}\label{boxsec}

Let us start by reviewing the general prescription to compute the index of theories of class ${\cal S}$.
The superconformal index~\cite{Kinney:2005ej,Romelsberger:2005eg} of an ${\cal N}=2$ SCFT can be thought of as a trace over states of the theory in the radial quantization, i.e.
a partition function on $S^3\times S^1$,
\begin{equation}\label{inddef}
{\cal I}=\mathrm{Tr} (-1)^F\,\left(\frac{t}{pq}\right)^r\,
 p^{j_{12}}\,
 q^{j_{34}}\,
 t^{R}\,
 \prod_i a_i^{f_i} \,.
\end{equation}
We denoted as $j_{12}$ as $j_{34}$ the rotation generators in two orthogonal planes:
$j_{12}=j_2+j_1$ and $j_{34}=j_2-j_1$ with $j_{1,2}$ being the Cartans of the Lorentz $SU(2)_1\times SU(2)_2$ isometry of $S^3$.
 $r$ is the $U(1)_r$ generator, and $R$ the $SU(2)_R$ generator of R-symmetries. 
The $a_i$ are fugacities for the flavor symmetry generators $f_i$.
The states which contribute to the index above satisfy
\be\label{BPS}
E-2j_2-2R+r=0\,.
\ee We have chosen to compute the index with respect to supercharge ${\widetilde{\cal Q}_{1\dot{-}}}$ which has the following charges:
$j_1=0$, $j_2=-\half$, $R=\half$, and $r=-\half$. All other choices of supercharges give equivalent results.

The  general prescription to compute the index of theories of class ${\cal S}$ corresponding 
to three-punctured spheres with punctures defined by auxiliary Young diagrams $\Lambda_\ell$ takes the following form~\cite{Gadde:2011uv}
\be\label{ind3pt}
{\cal I}={\cal N}_N\,\prod_{\ell=1}^3\hat {\cal K}(\Lambda'_\ell(a_\ell)) \;\sum_{\lambda} \frac{1}{\psi_\lambda(t^{\frac{1-N}{2}},\dots,t^{\frac{N-1}{2}})}
\prod_{\ell=1}^3\psi_\lambda(\Lambda_\ell(a_\ell))\,.
\ee
The sum over $\lambda\equiv(\lambda_1,\dots,\lambda_{N-1},0)$ is a sum over Young
diagrams with at most $N-1$ rows, i.e. over irreducible finite representations of $SU(N)$. 
 The prefactors $\hat K$ for the maximal punctures are given by elliptic Gamma functions~\cite{Gadde:2011uv,GRR}\footnote{
The abundant relevance of elliptic Gamma functions~\cite{Spiridonov4}
to the superconformal index computations was observed in~\cite{Dolan:2008qi}.}
\be
\hat K(a_1,\dots,a_N)=\prod_{i\neq j}\prod_{m,n=0}^\infty\frac{1-\frac{p q}{t} p^{m}q^{n}a_i/a_j}{1- p^{m}q^{n}t\,a_j/a_i}\equiv
\prod_{i\neq j}\Gamma\left(t\, a_i/a_j;\,p,\,q\right)\,.
\ee The functions $\psi_\lambda$ are given by an orthonormal set of eigenfunctions of the elliptic Ruijsenaars-Schneider model~\cite{GRR}.
The measure under which these functions are orthogonal is given by
\be
\hat \Delta=\frac{1}{N!}\prod_{i\neq j}\frac{\Gamma\left(t\, a_i/a_j;\,p,\,q\right)}{\Gamma\left( a_i/a_j;\,p,\,q\right)}\,.
\ee
In the limit $p=0$ (or $q=0$) this reduces to Macdonald measure and the functions $\psi_\lambda$ become Macdonald polynomials~\cite{Gadde:2011uv}.
For the discussion of this paper it will be sufficient to consider only the Hall-Littlewood (HL) limit of the index, $p=q=0$.  The HL index for \textit{linear} quivers is actually equivalent~\cite{Gadde:2011uv} to 
counting of chiral operators modulo superpotential constraints~\cite{Benvenuti:2010pq}, i.e. to the Hilbert series of the Higgs branch.\footnote{
A possible relation of a similar limit of the ${\mathcal N}=1$ index with the
counting problems discussed in~\cite{Gray:2008yu,Hanany:2008kn} was mentioned in~\cite{Spiridonov:2009za}.
} In principle,  the superconformal index is
not sensible to the superpotential constraints. However, in the case of linear quivers
 (for Lagrangian theories
and theories connected to those by dualities) the contributions of the
constraints to the Hilbert series precisely match the contributions of the fermions to the HL index.
For quivers corresponding to higher genus surfaces this is no more true: in general there will be
more fermions than constraints. We refer the reader to~\cite{Gadde:2011uv} for a detailed discussion of this
issue and in appendix~\ref{secapp} we give a couple of simple examples of relations between
Hilbert series, HL index, and Higgs branch countings.

\begin{figure}[htbp]
\begin{center}
\includegraphics[scale=0.6]{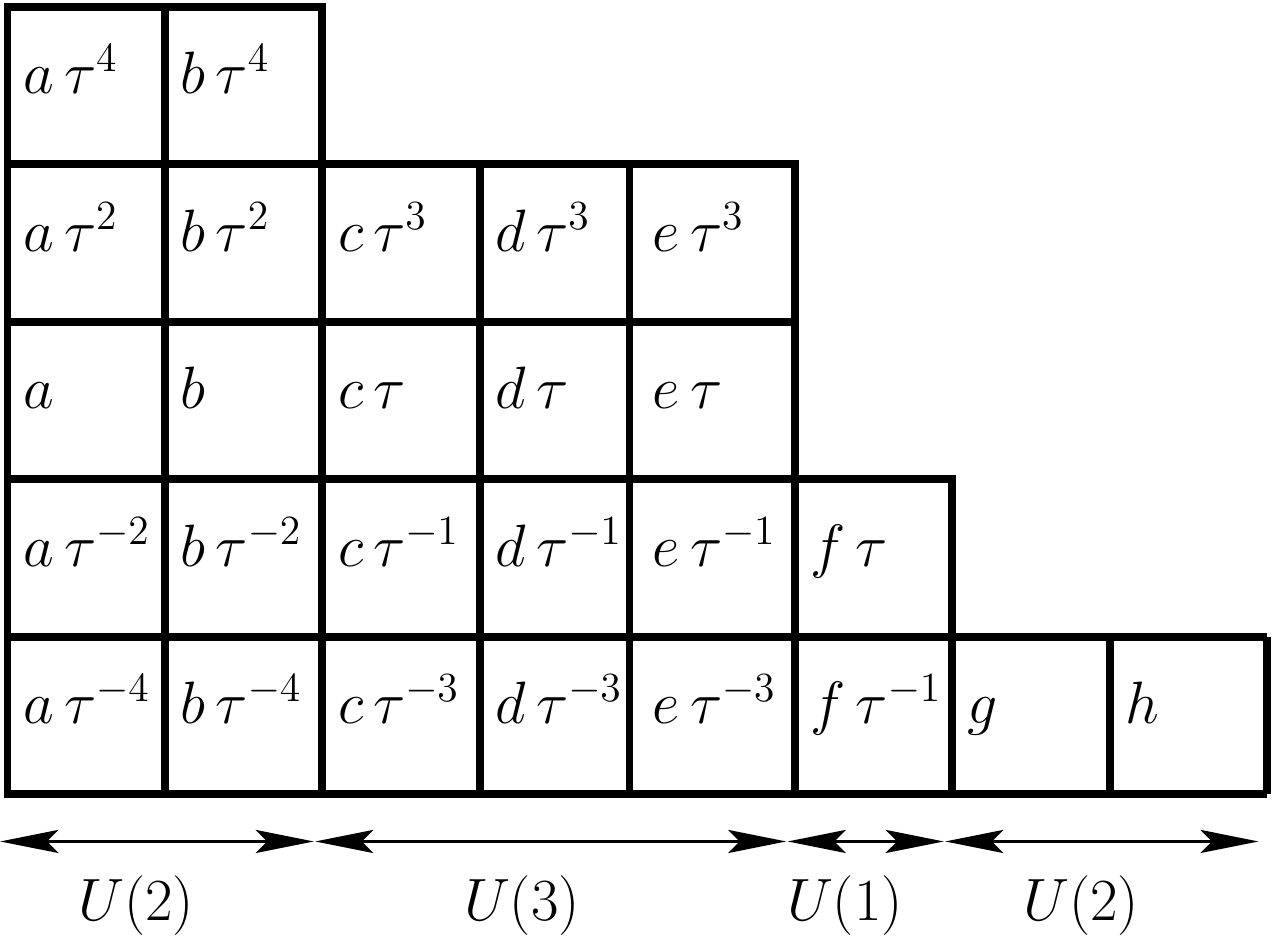}
\end{center}
\caption{Association of the flavor fugacities for a generic puncture. Punctures are classified
 by embeddings of  $SU(2)$ in $SU(k)$, so they are specified by the decomposition
 of the fundamental representation of $SU(k)$ into irreps of $SU(2)$, that is, by a partition of $k$.
 Graphically we represent the partition by an auxiliary Young diagram $\Lambda$ with $k$ boxes,
 read from left to right. In the figure we have the fundamental of $SU(26)$ decomposed as $\mathbf{5} + \mathbf{5} + \mathbf{4} + \mathbf{4} + \mathbf{4}  + \mathbf{2} + \mathbf{1} + \mathbf{1}$. 
 The commutant of the
embedding gives the residual flavor symmetry, in this case $S(U(3)\times U(2)\times U(2)\times U(1))$,
where the $S(\dots)$ constraint amounts to removing the overall $U(1)$.
 The $\t$ variable is viewed here as an $SU(2)$ fugacity, while the Latin variables are fugacities of the residual flavor symmetry.
The $S(\dots)$ constraint implies that the flavor fugacities satisfy
 $(ab)^{5}(cde)^4f^2gh=1$.\label{flavfig}}
\end{figure}

For the HL index  we can explicitly write down the eigenfunctions for $SU(N)$ theories as,
\be\label{HLdef}
\psi^{\lambda}(x_1,\dots,x_N|\;\t)=
{\mathcal N}_\lambda(\t)\;\sum_{\sigma \in S_N}
x_{\sigma(1)}^{\lambda_1} \dots x_{\sigma(N)}^{\lambda_N}
\prod_{i<j}   \frac{  x_{\sigma(i)}-\t^2 x_{\sigma(j)} } {x_{\sigma(i)}-x_{\sigma(j)}}\,,
\ee 
which are orthonormal
under the  measure
\be
\Delta_{HL}=\frac{1}{N!}\,\prod_{i\neq j}\frac{1-x_i/x_j}{1-\t^2 \,x_i/x_j}\, .
\ee
To avoid square roots in what follows we define
\be
\t=t^\half\,.
\ee
The normalization ${\mathcal N}_\lambda(\t)$ 
is given by
\be\label{normHL1}
{\mathcal N}^{-2}_{\lambda_1,...\lambda_N}(\t)=\prod_{i=0}^\infty \prod_{j=1}^{m(i)}\,
 \left(\frac{1-\t^{2j}}{1-\t^2}\right)\, ,
\ee where $m(i)$ is the number of rows in the Young diagram $\lambda=(\lambda_1,\dots,\lambda_N)$ of length $i$.
For $SU(N)$ groups we take Young diagrams with $\lambda_N=0$ and the product of $x_i$ in~\eqref{HLdef} is constrained as $\prod_{i=1}^Nx_i=1$.
Finally, the over-all normalization constant in~\eqref{ind3pt} in the HL limit is given by
\be
{\cal N}_N=(1-\t^2)^{2+N}\,\prod_{j=2}^N(1-\t^{2j})\,.
\ee
 The association of flavor fugacities $\Lambda_\ell(a)$ is illustrated in figure~\ref{flavfig}.
The functions $\psi_\lambda$ take $N$ arguments. If the flavor symmetry is smaller than $SU(N)$ then 
the $N$ arguments are read off each box in the auxiliary Young diagram defining the puncture.
We associate a flavor fugacity $a_i$ to each column of the auxiliary Young diagram defining the puncture.
Furthermore, in each column of height $k$ the boxes are assigned a factor of $\t^i$ ($i=k-1,k-3,\cdots 1-k$). 

\begin{figure}[htbp]
\begin{center}
\includegraphics[scale=0.6]{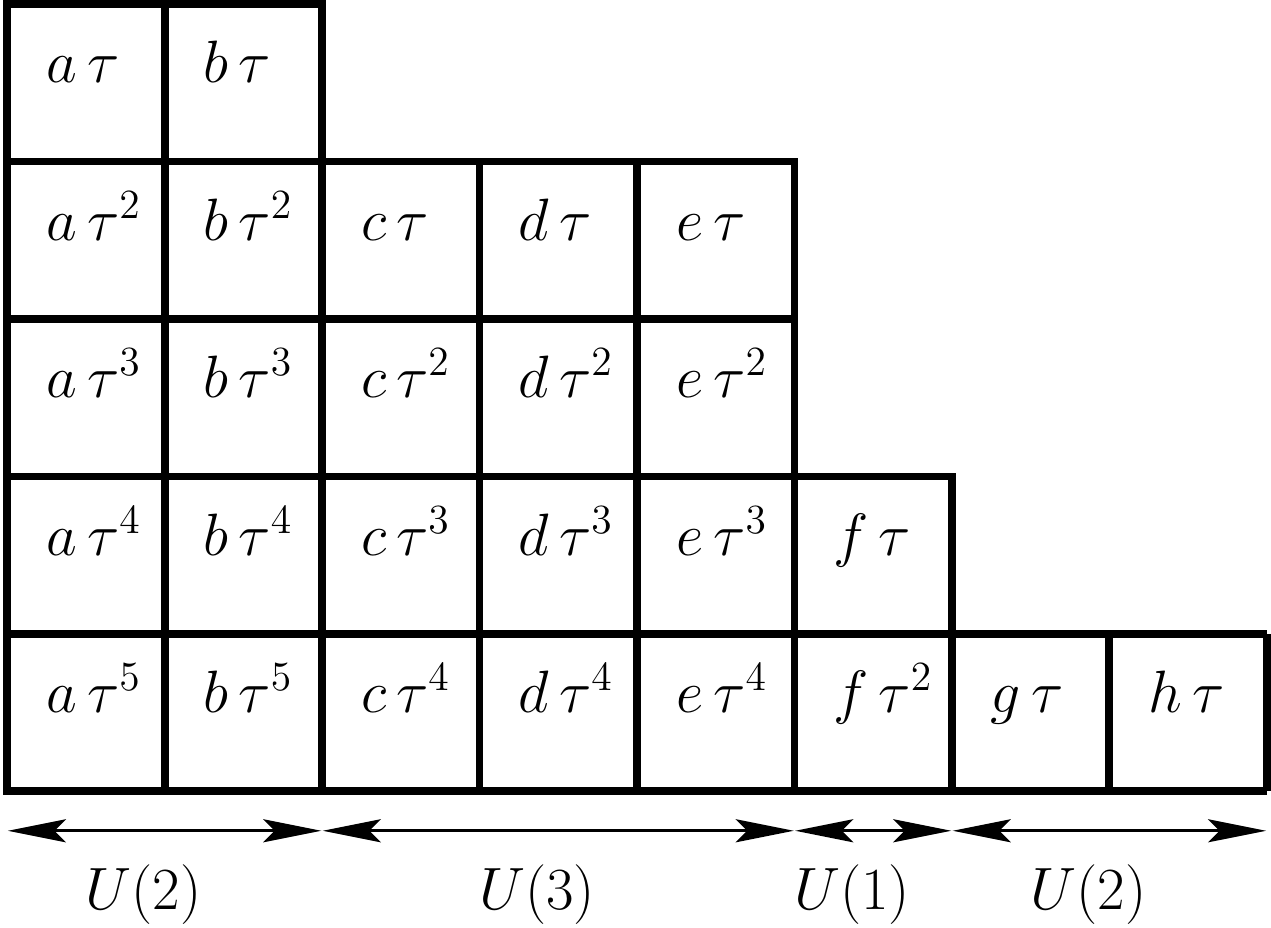}
\end{center}
\caption{The factors ${\frak a}^i_k$ associated to a generic Young diagram. The upper index is 
the row index and the lower is the column index. In $\bar{\frak a}^i_k$ one takes the inverse of flavor fugacities
while $\t$ is treated as real number. As before, the flavor fugacities in this example 
satisfy  $(ab)^{5}(cde)^4f^2gh=1$.\label{vectfig}}
\end{figure}

Similarly, the association of flavor fugacities $\Lambda'_\ell(a)$ is illustrated in figure~\ref{vectfig}.
The difference between the two functions is in the powers of $\t$.
The factors $\hat {\mathcal K}(\Lambda'(\mathbf{ a}))$ are defined as
\be\label{Ks} 
\hat {\mathcal K}(\Lambda'(\mathbf{ a}))=
\prod_{i=1}^{row(\Lambda)}\prod_{j, k=1}^{l_i}\frac{1}{1-{\frak a}^i_j\bar {\frak a}^i_k}\,.
\ee Here $row(\Lambda)$ is the number of rows in $\Lambda$ and $l_i$ is the length of $i$th row. The coefficients
${\frak a}^i_k$ are assigned as in figure~\ref{vectfig}.

For HL case the (inverse) structure constants (or the HL quantum dimension) has a simple and explicit form
\be
dim_\t(\lambda)\equiv \psi_\lambda(\t^{N-1},\t^{N-3},\dots,\t^{1-N}|\t)=
{\mathcal N}_\lambda(\t)\,\t^{\sum_{i=1}^{N-1}\left(2i-N-1\right)\lambda_i}\,
\prod_{i=1}^{N}\frac{1-\t^{2i}}{1-\t^2}\,.
\ee
Let us finally quote  the HL index of the genus ${\frak g}$ theory with $s$ punctures,
\be\label{gengs}
{\mathcal I}_{{\frak g},s} (\mathbf{a}_I; \t)=
(1-\t^2)^{(N-1)(1-{\frak g})+s}\,
\prod_{j=2}^{N}(1-\t^{2j})^{2{\frak g}-2+s}\,
\sum_{\lambda}
\frac{\prod_{i=1}^s \hat {\mathcal K}(\Lambda'_i({\mathbf a_i}))\;\psi^\lambda(\Lambda_i({\mathbf a_i})|\t)}
{\left[\psi^\lambda(\t^{{N-1}},\t^{{N-3}},\dots,\t^{{1-N}}|\t)\right]^{2{\frak g}-2+s}}\,.\nonumber\\
\ee
Several examples of using this prescription can be found in~\cite{Gadde:2011uv}.
The sum over the representations in~\eqref{gengs} can be written as a sum over a finite number
of geometric progressions. Thus the HL index can be always written as a ratio of two polynomials.
However, in practice these polynomials are of high degree and technically hard to evaluate.

\

\subsection{The Good, the Bad, and the Ugly: diagnostics}\label{trio}

The prescription of the previous section assumes that  the sum over representations
in~\eqref{gengs} converges. This is true in a vast majority of examples where the
underlying physical theories are known to exist. However, there are examples where it
is widely believed that the $4d$ theory exists but the prescription breaks down. Our 
purpose will be to understand better these cases and how to adjust the prescription to capture \textit{all}
the physically interesting situations. 

Let us seek a diagnostic tool to indicate when the sum diverges. It is easy to estimate the leading power 
of $\t$ in the expansion over the representations. Given a genus $\frak g$ $A_{N-1}$ theory with $s$ punctures of arbitrary types
and a representation of $SU(N)$ parametrized by $\lambda=(\lambda_1,\cdots,\lambda_{N-1},0)$,
where we always assume $\lambda_i\geq \lambda_{i+1}$, the leading power coming from the 
HL quantum dimension is 
\be
\t^{(2{\frak g}-2+s)\sum_{i=1}^N(N+1-2i)\lambda_i}\,.
\ee 
The leading power from the HL polynomials is given by
\be
\t^{\sum_{i=1}^N(\sum_{\ell=1}^sd_i^{(\ell)})\lambda_i}\,,
\ee where $d_{i}^{(\ell)}$ are the $N$ different powers of $\t$ for $\ell$th puncture 
which one can read from the auxiliary Young diagram as in figure~\ref{flavfig}. Moreover these numbers are ordered here
\be
d_{i}^{(\ell)} \leq d_{i+1}^{(\ell)}\,.
\ee 
For example, the puncture depicted in figure~\ref{flavfig} gives
\be
d=\left(-4,-4,-3,-3,-3,-2,-2,-1,-1,-1,-1,0,0,0,0,1,1,1,1,2,2,3,3,3,4,4\right)\,.
\ee
Thus, we can define a vector fixed by the theory at hand
\be
\upsilon_i\equiv (2{\frak g}-2+s)\,(N+1-2i)+\sum_{\ell=1}^s d_i^{(\ell)}\,,
\ee and the divergences appear if there exists a non-zero vector $\lambda$ such that
\be\label{bad}
\upsilon\cdot\lambda\leq 0\,.
\ee Note that if there is one such vector $\lambda$ automatically there is an infinite number of those by  rescaling $\lambda$ with positive integers.
We will call theory satisfying~\eqref{bad} a \textit{``bad''} one.
As a consistency check note that the contribution of no-puncture (single column) is zero,
\be
\delta\upsilon_i=N+1-2i+(2(i-1)-N+1)=0\,.
\ee
Let us give several examples. First, let us consider the extreme case of the torus without punctures for $A_{N-1}$ case. Here the vector $\upsilon$ is given by
\be
\upsilon=0\,,
\ee and the above condition is automatically satisfied for all representations. Here our prescription
fails because in the 2d TQFT language of~\cite{Gadde:2009kb} the torus with no punctures is a sum over states
in the Hilbert space with a unit weight and such a sum here is infinite.
However the index of ${\cal N}=2^*$ SYM corresponding to the torus with one simple puncture is finite since we have the following vector
\be
i=1\dots N-1:&&\qquad\upsilon_i=N+1-2i+(2(i-1)-N+2)=1\,,\\
i=N:&&\qquad \upsilon_i=1-N\,.\nonumber
\ee Now since $\lambda_N=0$ and since the numbers $\lambda_i$ are non-negative it is obvious that $\upsilon\cdot \lambda>0$.
This index is thus  converging.
Note that for representation $\lambda=(1,0,0,\dots)$ we have $\upsilon\cdot \lambda=1$: this will become important momentarily.

Let us consider trinions of $SU(3)$. First, the $SU(3)$ theory corresponding to a sphere with one maximal and two minimal punctures. We get
\be
\upsilon=(0,0,0)\,.
\ee Here it is obvious that any representation  gives $\upsilon\cdot \lambda=0$ and the index is 
divergent. There is no physical theory corresponding to this surface. Next we consider the free hypermultiplet: sphere with two maximal and one minimal punctures. Here we have,
\be
\upsilon=(1,0,-1)\,.
\ee Since $\lambda_3=0$ our criterion gives $\upsilon\cdot\lambda>0$ and the index is convergent. Here also for the choice $\lambda=(1,0,0)$ we get $\upsilon\cdot \lambda=1$.
Finally the trinion with three maximal punctures gives
\be
\upsilon=(2,0,-2)\,.
\ee Here also obviously $\upsilon\cdot\lambda>0$. Moreover, we do not have any representation giving  $\upsilon\cdot\lambda=1$. However, we
have  $\upsilon\cdot\lambda=2$ for $\lambda=(1,0,0)$: this will be also given a meaning shortly.
 
\

The next item in the trichotomy we want to define are the \textit{``ugly''} theories. These theories are defined as not bad theories which have 
a representation for which 
\be
\upsilon\cdot \lambda=1.
\ee
 The physical interpretation is as follows.
The HL index gets contributions only from states satisfying~\cite{Gadde:2011uv}
\be\label{HLconds}
E=2R+r\,,\qquad j_1=0\,,\qquad j_2=r\,.
\ee 
 The above condition implies that there is a bosonic state
contributing to the index with weight $\t^{2(E-R)=1}$. Thus this state has to have 
 $j_1=0$, $j_2=r$, $E=1-r$, and $R=\half-r$. Further, unitarity implies\footnote{
Averaging all the possible $\{Q,Q^\dagger\}$ operators  (see~\cite{Gadde:2011uv}
for the list of those) in this state and demanding positivity of the result has this implication.} 
 that 
the only possible choice is $r=0$: this is a scalar from the free hyper-multiplet.
This means that the theory splits into free components and (possibly)
interacting ones. We saw several examples above having this property. The ${\cal N}=2^*$ SYM is ${\cal N}=4$ SYM with decoupled free hyper, and the free hypermultiplet itself.
The trinion with maximal punctures however did not have such a term and indeed it does not have free components: it is the interacting $E_6$ SCFT~\cite{Minahan:1996fg}.
Theories which are not bad nor ugly will be called \textit{``good''} theories.

\

Finally we have the following additional diagnostic. If the theory is good and if there is a representation such that\footnote{Note that since the ugly theories have decoupled hypermultiplets the flavor symmetry
is automatically enhanced since we can rotate the half-hypermultiplets independently 
of the rest of the theory.}
\be
\upsilon\cdot \lambda=2,
\ee we expect that the flavor symmetry of the model will be enhanced. The physical motivation for this is that we expect all the operators 
contributing at $\t^2$ order to correspond to moment map operators of flavor symmetries.
From~\ref{HLconds} we get that these states have the following charges: $j_i=0$, $E=2$, $r=0$, and $R$=1.

 All moment map operators for the naive flavor symmetry are accounted for 
from
the $\hat {\cal K}$ factors in~\eqref{ind3pt} and from the sub-leading term in the singlet representation
in the sum ($\lambda=0$);  thus any state coming from the sum over non-zero representations enhances the flavor symmetry.
If we have an auxiliary Young diagram corresponding to flavor symmetry ${\cal G}=S\left(U(N_1)\times\dots U(N_\ell)\right)$ then the $\hat K$ factors contribute the following 
term at $\t^2$ order,
\be
\sum_{i=1}^\ell\sum_{k,j=1}^{N_i}a^{(i)}_j/a^{(i)}_k\,,
\ee where $a^{(i)}_j$ are the Cartans of $U(N_i)$. 
The contributions from the HL quantum dimension and  from the HL polynomial corresponding to the singlet representation $\lambda=0$ are
\be
\left[\dim_\t(0)\right]^{2-2\frak g-s}\to (1-\frak g)(N-1)-\half s (N-1)
\,,\qquad \psi_0^s\to\half s (N-1)\,.
\ee
Finally, the over-all normalization factor in~\eqref{gengs} gives $(N-1)(\frak g-1)-s$. Combining all these together we get that at the $\t^2$ order 
the contribution to the index from overall factors and from the singlet representation is,
\be
\sum_{I=1}^s\left[\sum_{i=1}^{\ell_I}\left(\sum_{k,j=1}^{N^{(I)}_i}a^{(i)}_{(I)j}/a^{(i)}_{(I)k}\right)-1\right]\,\t^2\,.
\ee This is exactly the contribution from the moment map operators of the symmetry ${\cal G}$.

 As we saw above we found a non-singlet representation contributing at $\t^2$ order
for the trinion of $SU(3)$ and indeed the flavor symmetry enhances from $SU(3)^3$ to $E_6$. On the other hand the  trinion of $SU(4)$ with maximal punctures has
\be
\upsilon=(3,1,-1,-3)\,,
\ee  and the lowest power one can get here is $3$: indeed the $SU(4)^3$ symmetry here does not enhance. For the $SU(4)$ theory with two maximal punctures and one square 
auxiliary diagram corresponding to $SU(2)$ flavor symmetry we get
\be
\upsilon=(2,0,0,-2)\,.
\ee Again we get terms of order $\t^2$ and indeed the symmetry here is enhanced to $E_7$.

\begin{figure}
\begin{center}
\includegraphics[scale=0.35]{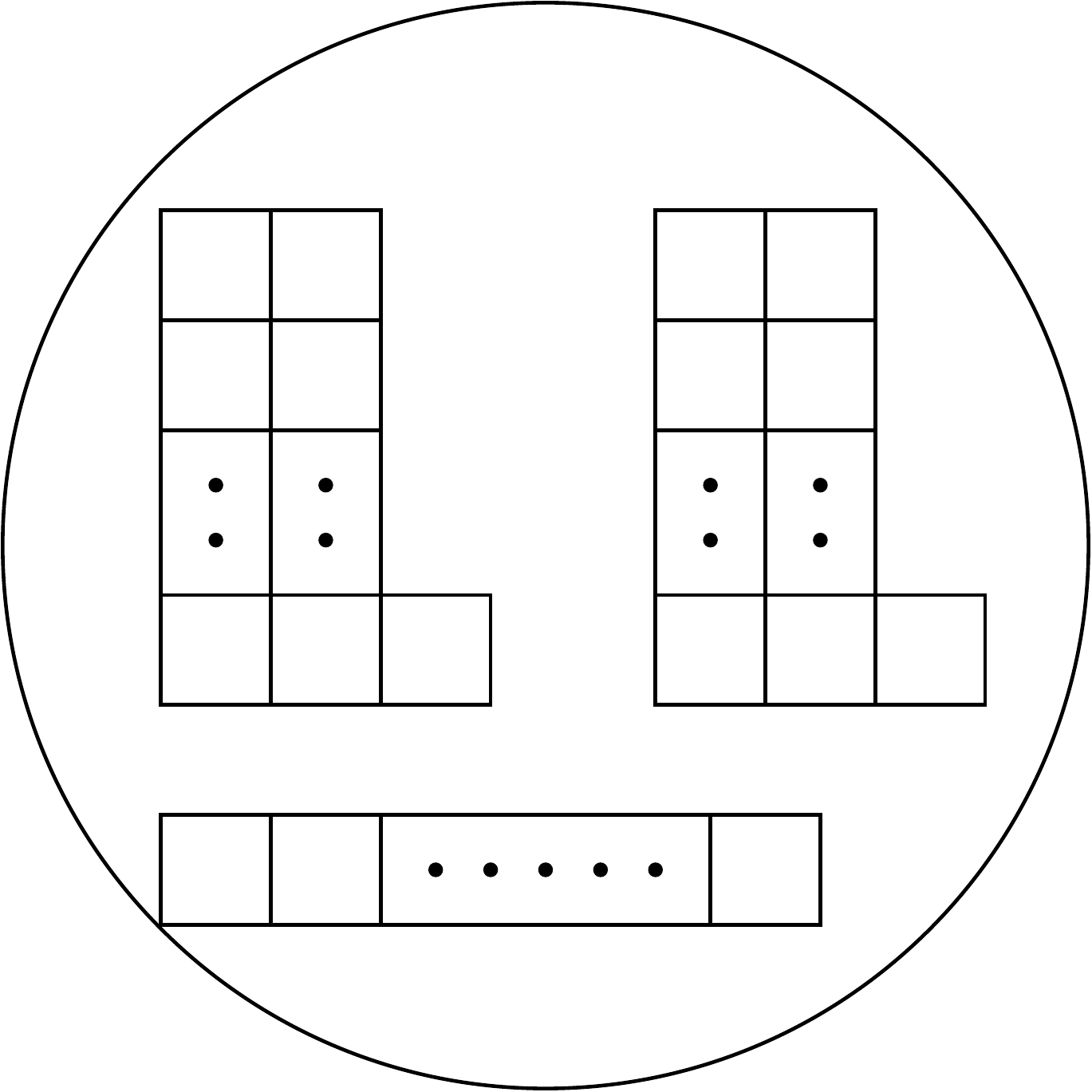}
\end{center}
\caption{The $SU(2k+1)$ Riemann surface corresponding to the rank $k$ interacting SCFT with flavor symmetry $SO(4k+6)\times U(1)$.
The height of the special punctures is $k$.\label{SOk}}
\end{figure}

Let us provide additional examples. We consider an $SU(5)$ theory corresponding to sphere with three punctures: one maximal and two special. The special punctures have two rows with two 
boxes in the first one and three boxes in the second one. The naive flavor group here is $SU(5)\times SU(2)^2\times U(1)^2$.
 Here we are getting 
\be
\upsilon=(2,0,0,0,-2)\,,
\ee and thus expect the flavor symmetry to be enhanced. Indeed to order $\t^2$ the index turns out to be
\be
{\cal I}=1+92\,\t^2+\cdots\,.
\ee The $92$ dimensional adjoint representation is actually an adjoint of $U(1)\times SO(14)$ (see the derivation of this from
S-dualities in~\cite{Chacaltana:2010ks}). 
In fact there is a whole series of theories corresponding to $SU(2k+1)$ quivers with three punctures: one maximal and two special ones 
as depicted in figure~\ref{SOk}. The theories of this sort were denoted $R_{2,k}$ in~\cite{Chacaltana:2010ks}. 
These theories are rank $k$ and the symmetry enhances to $SO(4k+6)\times U(1)$. In the case $k=1$ the $SO(10)\times U(1)$ enhances
further to $E_6$.
 Let us compute the $\upsilon$ here as another illustration of the procedure.
For the two special punctures we have 
\be
k \text{ odd}:&&\qquad d^{(\ell)}=(1-k,1-k,3-k,3-k,\cdots,-1,0,1,\cdots,k-1,k-1)\,,\\
k \text{ even}:&&\qquad d^{(\ell)}=(1-k,1-k,3-k,3-k,\cdots,0,0,0,\cdots,k-1,k-1)\,.\nonumber
\ee For maximal puncture always $d=0$. Then we get
\be
k \text{ odd}:&&\qquad\upsilon=(2,0,2,0,2,0,\cdots,0,-2,0,-2)\,,\\
k \text{ even}:&&\qquad\upsilon=(2,0,2,0,2,0,\cdots,0,0,0\cdots,0,-2,0,-2)\,.\nonumber
\ee

Another example of $SU(5)$ theory with enhanced symmetry is three punctured sphere with two maximal punctures and an L-shaped one corresponding to symmetry $S(U(1)\times U(2))$. The naive symmetry 
is $SU(5)^2\times SU(2)\times U(1)$. However it is enhanced to $SU(10)\times SU(2)$. An example of higher genus theory with enhanced symmetry is genus two surface of $SU(2)$ without punctures,
\be
\upsilon=(2,-2)\,.
\ee Here the index is a polynomial
\be
{\cal I}=1+\t^2-\t^4\,.
\ee We have a $U(1)$ flavor symmetry
giving opposite charges to the two hypermultiplets.
This index can not be interpreted as a Higgs branch Hilbert series since it is
associated to higher genus surface. Indeed, the $\t^4$ term comes with negative sign 
indicating that there are more fermions contributing to the index than there are
superpotential  constraints.
 For higher rank genus two surfaces without punctures there is no
enhancement of flavor symmetry,
\be
\upsilon=(2N-2,2N-4,\cdots,2-2N) \,.
\ee	
Taking $A_2$ torus with one maximal puncture we have
\be
\upsilon=(2,0,-2)\,,
\ee and thus expect an enhancement of the $SU(3)$ symmetry: indeed it enhances to $G_2$.
  The index actually takes the following form
\be
{\cal I}_{G_2}&=&1+\chi^{G_2}_{14}(a)\t^2+\left(\chi^{G_2}_{Sym^2{\bf {14} }}(a)-\chi^{G_2}_7(a)-1\right)\t^4+\\
&&\qquad+\left(\chi^{G_2}_{Sym^3{\bf {14} }}(a)-(\chi^{G_2}_7(a)+1)\chi^{G_2}_{14}(a)\right)\t^6+
\cdots\,,\nonumber
\ee where
\be
\bf{14}_{G_2}=\bf{3}+\bf{\overline{3}}+\bf{8}\,,\qquad
\bf{7}_{G_2}=1+\bf{3}+\bf{\overline{3}}\,.
\ee Here we start with the $A_2$ trinion and glue together two legs, i.e. gauge a diagonal $SU(3)$:
the commutant subgroup is $G_2$.
The $E_6$ adjoint representation decomposes as
\be
\bf{ 78}=\bf{8}+\bf{8}_1+\bf{8}_2+\bf{3}\times\bf{3}_1\times\bf{3}_2+
\bf{\overline{3}}\times\bf{\overline{3}}_1\times\bf{\overline{3}}_2\,.
\ee In diagonal $SU(3)'$ we identify $SU(3)_1$ with the conjugate of $SU(3)_2$. Thus,
we get a decomposition into $G_2$ representations,
\be
\bf{ 78}=\bf{8}'+(\bf{8}+\bf{3}+\bf{\overline{3}})+\bf{8'}(\bf{3}+\bf{\overline{3}}+1)\,.
\ee
Note that this is again genus one quiver and thus a-priori we do not have an interpretation of the result as
a Hilbert series of the Higgs branch.

\

\subsection{Reducing flavor symmetry and residues: an example}

It has been shown in~\cite{GRR} that one can reduce the amount of flavor symmetry of a theory
by computing certain residues.  We will concentrate in what follows on the HL index, but the residue 
prescription we will discuss is valid for more refined versions of the index as well. Let us first give a simple example.

We will compute the index of an arbitrary $A_{N-1}$ theory with one L-shaped puncture: the auxiliary Young diagram 
consists of two rows with one box in the first row and $N-1$ boxes in the second. 
To do so we will start from the same theory with an L-shaped puncture traded with a maximal one
and compute a certain residue. Then we will compare to the prescription of~\cite{Gadde:2011uv} which we reviewed 
in the beginning of this section.

\begin{figure}
\begin{center}
\includegraphics[scale=0.45]{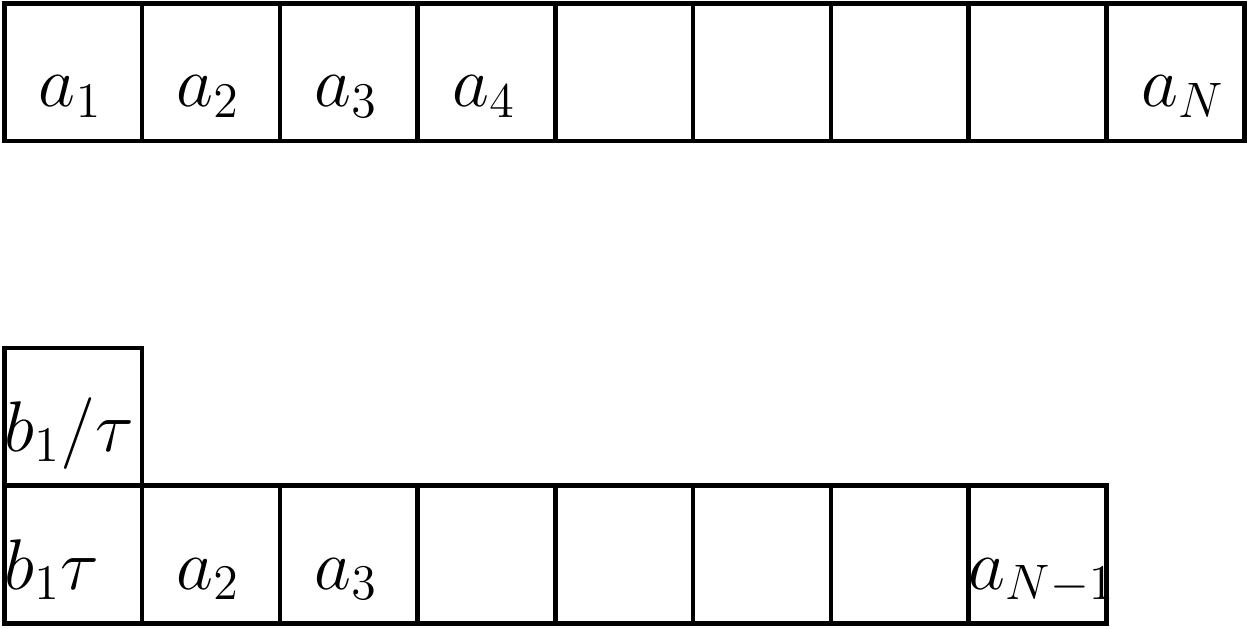}
\end{center}
\caption{Getting L-shaped puncture from a maximal one.\label{LL}}
\end{figure}

Let us start with a maximal puncture.
The factor $\hat{\cal K}$ for the maximal puncture has the following form
\be
\hat {\cal K}
=\prod_{i,j}\frac{1}{1-\t^2\,a_i/a_j}\,.
\ee This expression has poles whenever
\be
a_i=\t^2\, a_j\,.
\ee Since $\prod_{i=1}^Na_i=1$ we can also write
\be
a_i=\t\, \frac{1}{\prod_{j\neq i}^{N-1}a_j^{1/2}}\,.
\ee We will treat the fugacities $a_{1,\dots,N-1}$ as independent.
 Let us consider the pole at $a_1$ and define $b_1=\prod_{j\neq 1}^{N-1}a_j^{-1/2}$.
We put $a_1$ at the left-most edge of the single row corresponding to the maximal puncture (see figure~\ref{LL}).
Then the fugacity assignment in the orthogonal functions, which are HL polynomials in this paper and thus do not have poles themselves,
 will be
\be
\psi^\lambda(a_1,\cdots,a_N)\quad\to\quad
\psi^\lambda(\t b_1,\t^{-1}b_1,a_2\cdots,a_{N-1})\,,
\ee  and of course by construction $b_1^2\prod_{i=2}^{N-1}a_i=1$. 
In the prescription of~\cite{Gadde:2011uv} the above assignment corresponds to two row Young diagram with one box
in the first row and $N-1$ boxes in the second and defining a puncture with flavor symmetry
$S(U(1)\times U(N-2))$.
Let us now compute the residue of the index at this pole\footnote{We use the short notation
implying that when a term with ambiguous signs appears we should consider a product
of that term with all possible choices of the sign.}
\be
&&\underset{a_1\to \t^{}b_1}{Res}\{\hat {\mathcal K}(a_i)\}=\\
&&\qquad\half\frac{1}{(1-\t^2)^2(1-\t^4)}\prod_{i,j=2}^{N-1}\frac{1}{(1-\t^2\,a_i/a_j)}
\prod_{i=2}^{N-1}\frac{1}{(1-\t^{3}(a_i/b_1)^{\pm1})(1-\t^{}(b_1/a_i)^{\pm1})}\,.\nonumber
\ee  
On the other hand the prescription of~\cite{Gadde:2011uv} tells us that the $\hat {\cal K}$ factor corresponding to the L-shaped puncture is 
\be 
\hat {\mathcal K}_{L}(a_i,b_1)=
\frac{1}{(1-\t^2)(1-\t^4)}\prod_{i,j=2}^{N-1}\frac{1}{(\t^2\,a_i/a_j)}
\prod_{i=2}^{N-1}\frac{1}{(1-\t^{3}(a_i/b_1)^{\pm1})}\,.
\ee The ratio of the two quantities above is simply given by
\be
\frac{\hat {\mathcal K}_{L}(a_i,b_1)}{\underset{a_1\to \t^{}b_1}{Res}\{\hat {\mathcal K}(a_i)\}}=2\;
{\cal I}_{V}\,
\prod_{i=2}^{N-1}(1-\t^{}(b_1/a_i)^{\pm1})\,.
\nonumber\\
\ee Here, ${\cal I}_V=1-\t^2$ is the index of free vector multiplet.
Note that the other factor on the right-hand side is an inverse of the index of a free hypermultiplet
in representation $\bf{(N-2)}_-+\overline {\bf{(N-2)}}_+$ of  $S(U(1)\times U(N-2))$.
Thus the prescription to obtain the index of a theory with $S(U(1)\times U(N-2))$ 
puncture from the index of the theory with the maximal puncture is just to consider the pole
described above and multiply the index by the index of a free vector multiplet 
and divide by the index of  an appropriate free hyper-multiplet. 
 This procedure is generalizable 
to arbitrary punctures by iterating the above and we will explicitly show this in the next section.

In this derivation we assumed that the sum over the representations 
converges when we consider the special assignment of flavor fugacities
and thus the only pole comes from the over-all factors. This fact need not be always true
as we saw in the previous subsection.
When too many punctures are closed too much this sum does not converge anymore
and a good (ugly) theory might become bad. We will discuss this issue in the next section.

\

\subsection{Lifting boxes}

Let us now state the general procedure to write the index of a theory with smaller flavor symmetry by a residue
computation of a theory with a bigger flavor symmetry.\footnote{The residue prescription here is given for the HL index but it can be generalizaed to the index
with more superconformal fugacities turned on.}
 Given the index, ${\cal I}_\Lambda$, of a theory with a puncture 
corresponding to a Young diagram $\Lambda$ with flavor symmetry ${\cal G}_\Lambda$
we can construct the index, ${\cal I}_{\Lambda'}$, of a theory with puncture $\Lambda'$ differing from $\Lambda$
by a position of one box and having smaller flavor symmetry,
\be\label{liftp}
{\cal I}_{\Lambda'}=(\ell+1)\frac{{\cal I}_{V}}{{\cal I}_{hyp}(\Lambda,\Lambda')}\;
\underset{a\to \t B}{Res}{\cal I}_\Lambda \,.
\ee
The quantity ${\cal I}_{hyp}(\Lambda,\Lambda')$ is the index of a free hypermultiplet in a representation
we will determine shortly. A reader not interested in the straightforward technical details 
of the derivation of the above formula can safely skip the rest of this subsection.

\

Since the residue prescription involves factoring out certain contributions the order in which
the procedure is done is essential. We will build the punctures starting from the maximal one
by filling up the left-most column first, then going to the next column, and so on.
 To do so we will raise the right-most box to the desired position.
The parameter $\ell$ in~\eqref{liftp} is the height of the column to which the box is lifted.
From its definition~\eqref{Ks}, the factor of $\hat K$ has a pole at the position,
\be
a^{\ell+1}=\t^{{\ell+1}}\, B^{\ell+1}\,.
\ee The fugacity $a$ is associated with the column we are lifting the box to. 
The factor $B^{\ell+1}$ is defined to be the inverse of the product of all the 
fugacities other than $a$.  If we consider the above mentioned pole 
then the right-most box, fugacity of which we take to depend on the others,
 receives the value $\t^{-{\ell}}\, B$, and the 
fugacity $a$ become $\t B$. Thus in the orthogonal polynomials
we obtain
\be
\psi^\lambda(\cdots,\t^{{\ell-1}}a,\t^{{\ell-3}}a,\dots,\t^{{1-\ell}}a,\frac{B^{\ell+1}}{a^{\ell}})\quad\to\quad
\psi^\lambda(\cdots,\t^{{\ell}}B,\t^{{\ell-1}}B,\dots,\t^{-{\ell}}B)\,,
\ee which exactly corresponds to the Young diagram with the box lifted to the desired position.
We will see in what follows that ${\cal I}_{hyp}(\Lambda,\Lambda')$ is equal to
\be\label{hypdec}
{\cal I}_{hyp}(\Lambda,\Lambda')=\frac1{1-{\t}{\mathbf a}\,b^{-1}}\frac1{1-{\t}{\mathbf a}^{-1}\,b^{}},\qquad b=\t^{{1-\ell}}\,B\,.
\ee Here ${\mathbf a}$ are the fugacities of the single row tail of $\Lambda'$. 
\begin{figure}
\begin{center}
\includegraphics[scale=0.5]{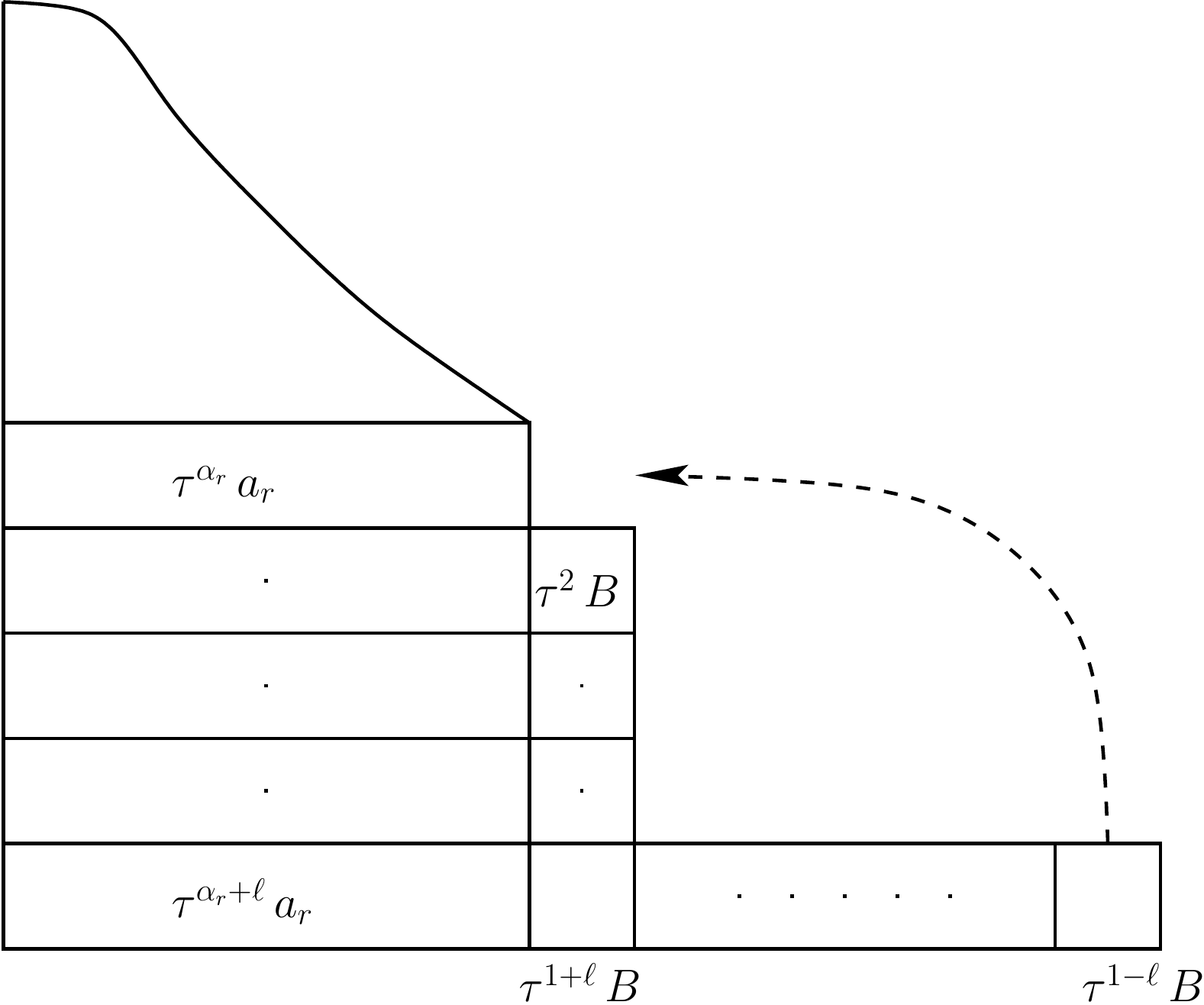}
\end{center}
\caption{Auxiliary Young diagram. The flavor symmetry is decreased by raising the right-most box 
to the position indicated by the arrow.\label{proof}}
\end{figure}
Let us derive~\eqref{hypdec}.
This factor is derived by comparing the residue of $\hat {\cal K}_\Lambda$ with $\hat {\cal K}_{\Lambda'}$.  
Factors of $\hat {\cal K}$ are obtained by going over each row of the Young diagram and summing over contributions
of each ordered pair of boxes, see~\eqref{Ks}. 
To compare the residue of $\hat {\cal K}_\Lambda$ with $\hat {\cal K}_{\Lambda'}$ it is useful to split the auxiliary Young diagram to three regions: (1)
the top of the diagram (the curly ``hat'' in figure~\ref{proof}), (2) the $\ell+1$ lowest rows without the tail (the columns of height one), and the tail (3). 
 The contributions of the rows in the  ``hat'' (1) of figure~\ref{proof} is the same for both Young diagrams. We only have to compare the $\ell+1$ lowest rows.
The contribution of one of the rows in region (2) results in a product of two different types of factors,
\be
\frac1{1-\t^{\a_r+2i-1}a_r/B}\,,\qquad
\frac1{1-\t^{\a_r+2i+1}B/a_r}\,.
\ee The latter is the same as the analogous factor in the same row in $\Lambda'$ but the former
actually corresponds to the previous row, $i-1$, in $\Lambda'$. 
Now the right-most box contracted with the boxes to the left of the column on top of which it is 
moved gives
\be
\frac1{1-\t^{\a_r+2\ell+1}a_r/B}\,,\qquad
\frac1{1-\t^{\a_r+1}B/a_r}\,.
\ee The former gives the missing factor for the lowest row and the latter
gives the missing factor for the row
where the extra box is moved to, $(\ell+1)$th row. The contractions of the right-most box
and the lowest box in the column to which we move the box  gives the pole residue of which we compute 
and
\be
\frac1{1-\t^{2\ell+2}}\,.
\ee
This
factor gives the correct self contraction of the lowest box in
this column in $\Lambda'$. The self-contraction of the right-most box 
gives the factor of  
\be
\frac1{1-\t^{2}}\,.
\ee  This factor is missing in the diagram $\Lambda'$ and we have to factor it out.
Finally, the contraction of the unit height tail (region (3)) with  the right-most box and with 
the bottom box in the column where we move to  yields
\be
\frac1{1-{\t^{2+\ell}}{\mathbf a}/B}\frac1{1-{\t^{2+\ell}}B/{\mathbf a}}\,,\qquad
\frac1{1-{\t^{}}{\mathbf a}/(B/\t^{\ell-1})}\frac1{1-{\t^{}}(B/\t^{\ell-1})/{\mathbf a}}\,.
\ee The former factor appears also in $\Lambda'$ but the latter has to be factored out.
To summarize, starting from diagram $\Lambda$ the residue prescription gives the 
diagram $\Lambda'$ and extra factors
\be
\frac1{1-\t^2}\,
\frac1{1-{\t^{}}{\mathbf a}/(B/\t^{\ell-1})}\frac1{1-{\t^{}}(B/\t^{\ell-1})/{\mathbf a}}\,.
\ee The first factor is just the inverse of the index of a free vector and the second 
is a free hyper in representation $\overline{\bf k}_{+1}+{\bf k}_{-1}$
 with the $U(1)$ charge $\pm1$ coupled to fugacity $B/\t^{\ell-1}$, and $k$ being the length of the tail of $\Lambda'$. 

Since the prescription involves taking a residue of $U(1)$ flavor fugacity and multiplying by the 
index of $U(1)$ vector field it is natural to interpret the procedure as computing the index
of the theory corresponding to diagram $\Lambda'$ by gauging a $U(1)$ flavor symmetry in theory
corresponding to diagram $\Lambda$. 
See~\cite{GRR} for a detailed discussion of this intuition. 

\

\section{Going beyond the ugly theories}\label{uglysec}

In the previous section we have formulated a prescription to compute the index of a generalized quiver theory 
by starting from the model with all maximal punctures and performing certain residue computation.
The poles we considered appear in pre-factors $\hat K$ associated with each puncture. 
The procedure of taking these residues presumes that the {\textit{whole}} index has the relevant simple pole.
This assumption can fail in two ways,
\begin{itemize}
 \item The sum over representations has a zero at the relevant value of the flavor fugacity and the pole disappears.
\item The sum over representations has a pole at the relevant value of the flavor fugacity and the simple pole
becomes higher order pole.
\end{itemize}
In the following we will see examples of these two situations and discuss the consequences.

\subsection{Disappearing poles}

An example of the former case is trying to compute the index of a putative theory
corresponding to a Riemann surface with one maximal and two minimal punctures.
The basic such example is the $A_2$ case: we will denote such a vertex here as $311$.
As we saw in section~\ref{trio} this is a bad theory and thus the prescription of~\cite{Gadde:2011uv} fails here.
 Following the residue prescription
of the previous section we obtain the index $311$ by starting from the index 
of $331$, the free hyper-multiplet, and computing the residue at the pole
$a_1=\t\,B$, where $B$ is a product of fugacities associated to the same 
puncture as $a_1$. The HL index of a free hypermultiplet is given by
\be
{\cal I}_{331}=\prod_{i,j=1}^3\frac{1}{1-\t{a_ib_j c}}\frac{1}{1-\t\frac{1}{a_ib_j c}} \,,
\qquad \prod_{i=1}^3 a_i=\prod_{j=1}^3b_j=1\,.
\ee However, this index \textit{does not} have the corresponding pole! What happens
is that the pole in $\hat {\cal K}$ is canceled out by a zero in the sum over representations.
In particular applying the residue prescription we get identically zero implying that there is no theory
corresponding to the $311$ vertex. 

This is consistent with the fact that $311$ vertex does not make sense as an independent entity.
In particular it is useful to compute it's rank. One can do so by using the Argyres-Seiberg duality~\cite{argyres-2007-0712}.
In one duality frame the theory corresponding to four-punctures sphere with two maximal and two minimal punctures
is just ${N}_f=6$ SYM and thus has rank $2$. In the other duality frame, where the two minimal punctures are taken to collide, 
the same theory can be thought of as the $E_6$ Minahan-Nemeschanski SCFT~\cite{Minahan:1996fg}, which has rank $1$, coupled through $SU(3)$ gauge field 
to $311$ vertex. This implies that the naive rank of $311$ vertex is $-1$.\footnote{It is interesting to note 
that our definition of a bad theory here relies on Higgs branch diagnostics. However,
in all the cases we considered the bad theories with sphere topology  
also had a negative ``naive'' rank~\cite{Gadde:2011uv} (as defined say below  equation (1) in~\cite{Chacaltana:2010ks}) which is
a Coulomb branch diagnostics.} 
Thus the vertex $311$ only makes sense as a part of bigger theory and does not appear by itself
as a sensible quantum field theory.  

\

\subsection{Higher order poles}

Let us give a couple of examples of cases when the order of the simple pole is increased.

\subsubsection*{From ${\cal N}=2^*$ SYM to ${\cal N}=4$ SYM}

The simplest case of a double pole is ${\cal N}=2^*$ SYM, i.e. the theory corresponding to torus with one 
minimal puncture. We can try to remove the single puncture of this theory by raising a single box. However, here 
the sum over representations also has a pole and the corresponding theory is bad.
 The index of the torus with one minimal puncture is given by (we give the example of $A_1$ model for simplicity)
\be
{\cal I}_{{\cal N}=2^*}=\frac{1}{1-\t a^{\pm1}}\,\frac{1+\t^2-\t^3(a+a^{-1})}{1-\t^2a^{\pm2}}=\frac{1}{1-\t a^{\pm1}}\,\;{\cal I}_{{\cal N}=4}\,,
\ee and indeed this has a second order pole at $a=\t$. The prescription of~\cite{Gadde:2011uv} to compute the index of the torus thus diverges in this case.
Moreover the residue prescription, though giving a finite result, does not make sense physically.
The physical reason for this can be traced to the following.  The ${\cal N}=2^*$ SYM is actually equivalent to 
${\cal N}=4$ SYM with a decoupled hypermultiplet, i.e. it is an ugly theory. Since the free hypermultiplet and the interacting ${\cal N}=4$ SYM are decoupled
there is an enhancement of symmetry and it physically actually makes sense to define different flavor fugacity $a$ for the two components,
\be
{\cal I}_{{\cal N}=2^*}=\frac{1}{1-\t a_1^{\pm1}}\,\frac{1+\t^2-\t^3(a_2+a_2^{-1})}{1-\t^2a_2^{\pm2}}\,.
\ee Here, $a_1$ couples to the $sp(1)$ symmetry rotating the two half-hypers of the free component and $a_2$ couples
to a combination of R-symmetries of the SYM. Thus, after this refinement considering the pole in $a_1=\t$ we recover our prescription.\footnote{
The residue at $a_1=\tau$ gives ${\cal I}_V^{-1}$ times the index of ${\mathcal N}=4$ SYM. Thus, as instructed by~\eqref{liftp} we strip off the index of the free vector. 
The Young diagram corresponding to no-puncture does not have a ``tail'' and thus there is no hypermultiplet to be stripped off. 
A naive application of~\eqref{liftp} will also imply that we have to multiply the index by a factor of $2$. We do  not have to do so here though: as is discussed in~\cite{GRR}
such numeric factors are usually associated to ``gauging'' discrete symmetries, i.e. to having poles also at $a=\exp[\frac{2\pi i\,k}{\ell+1}]\t$
for integer values of $k$; which we do not have here.  
}
The ${\cal N}=2^*$ is an example of an ``ugly'' theory since it has decoupled free components and this is exactly the reason why the
naive computation of the index is obstructed.
\begin{figure}
\begin{center}
\includegraphics[scale=0.35]{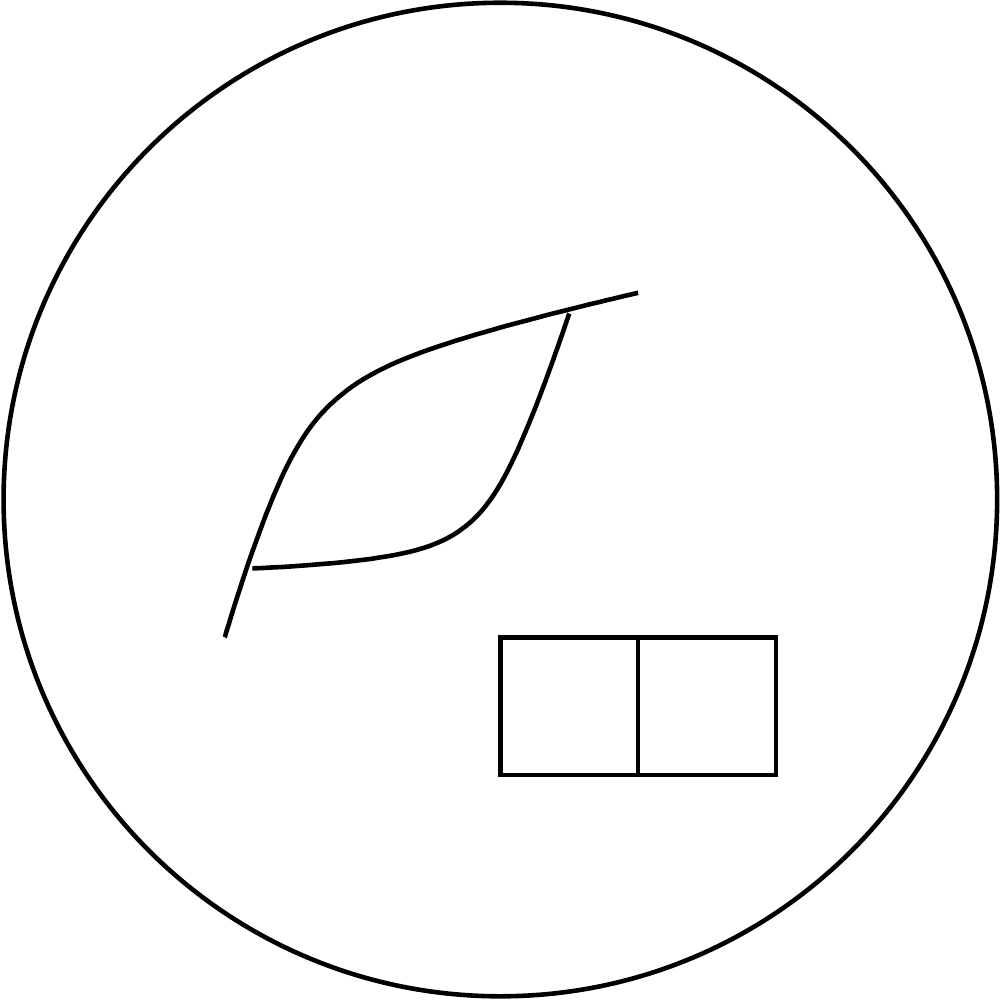}
\end{center}
\caption{The Riemann surface corresponding to the ${\cal N}=2^*$ theory. The torus without punctures,
i.e. the ${\cal N}=4$ SYM is obtained by lifting the right-most box
on top of the left-most one.\label{starfig}}
\end{figure}

\

\subsubsection*{From $A_3$ quiver to SYM with $usp(4)$ gauge group}

Let us study now another case of a theory with a double pole by considering the example
of four punctured sphere of $SU(4)$ type with three square 
punctures with flavor symmetry $SU(2)$ and one $L$ shaped puncture
corresponding to $S(U(1)\times U(2))$ flavor symmetry (let us denote this theory by $222L$).
This is an ugly theory of rank two with  finite index,
\be
\upsilon=(2,-1,1,-2)\,.
\ee The representation $\lambda=(1,1,0,0)$ gives contribution at order $\t$ and thus we expect to have a free component.
We will further try to close the $L$ shaped
puncture to a square $SU(2)$ one; the theory with four square-shaped punctures
is a ``bad'' one, index of which naively diverges,
\be
\upsilon=(2,-2,2,-2)\,.
\ee The representations $\lambda=(\ell,\ell,0,0)$ all contribute at order $\t^0$. 

The association of the flavor fugacities to the $L$-shaped
puncture is $(\t^{-1} b,\t b, \frac{a}{b} ,\frac{1}{ab})$. The limit
to $2222$ theory, theory with four square punctures, corresponds to taking $a\to \t$.
Computing the index of $222L$ one can see that the sum over representations
has a simple pole in flavor fugacities in the above mentioned limit. Since a simple pole
is also present in an over-all factor $\hat K$ we have here a double pole. As we mentioned before the HL index is always a ratio of two polynomials.
Keeping only the flavor fugacity
with respect to which we compute the pole, $a$,
different from one the denominator of the HL index is 
\be
(a-\t )^2 (a+\t ) (-1+a \t )^2 (1+a \t ) \left(-1+\t ^2\right)^{12} \left(a-\t ^3\right)^{10} \left(-1+a \t ^3\right)^{10}\,.
\ee 
Here again,
the prescription 
of~\cite{Gadde:2011uv} for the $2222$ index diverges since this theory is a bad one, and the residue prescription gives finite
but physically non-sensical result.
\begin{figure}
\begin{center}
\includegraphics[scale=0.35]{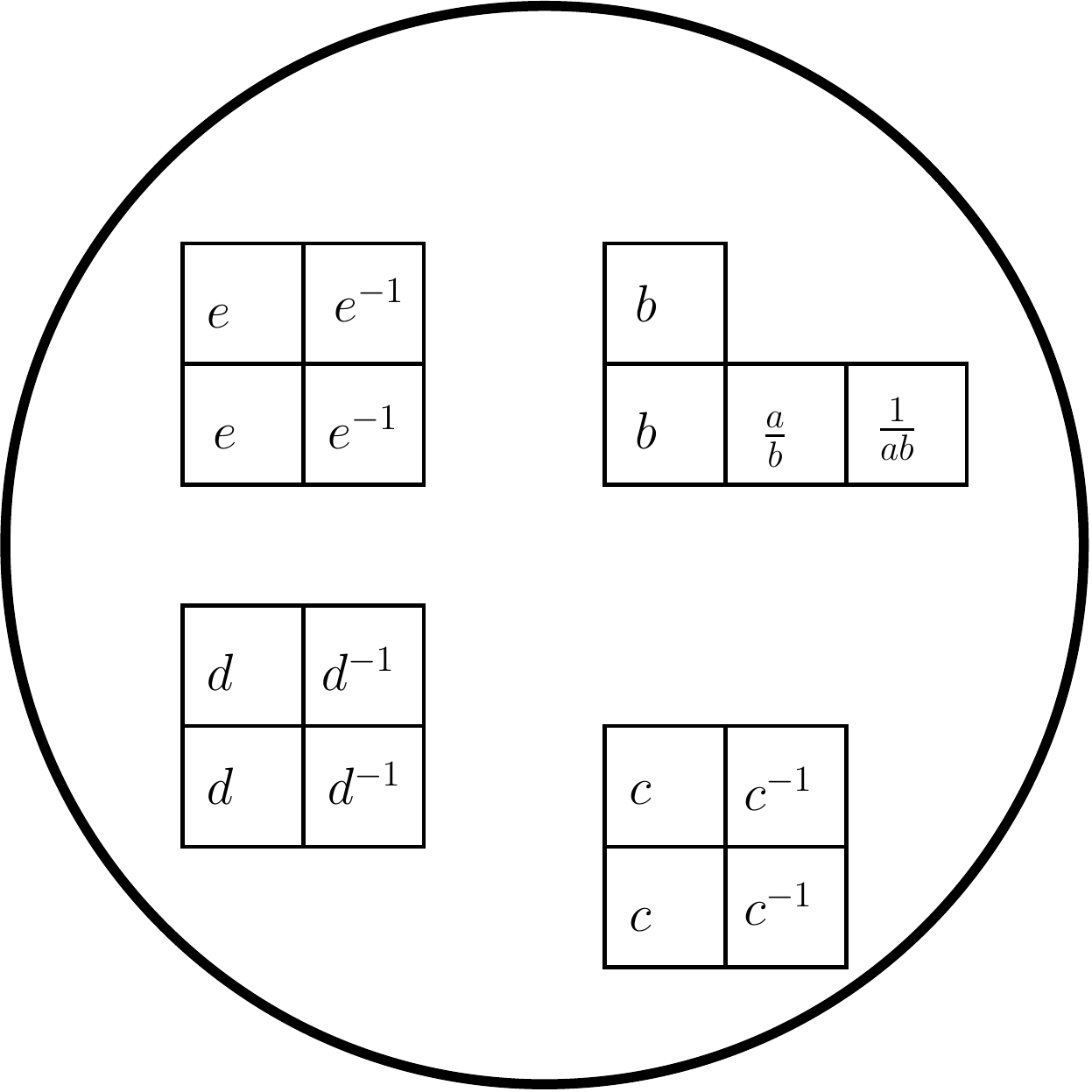}
\end{center}
\caption{Association of  flavor fugacities for the vertex corresponding to the $222L$ theory.}
\end{figure}

The $2222$ theory is believed~\cite{Benini:2009gi,Moore:2011ee} to be a $usp(4)$ gauge theory with four hypermultiplets in 
fundamental representation of the gauge group and one in the antisymmetric representation.
Actually computing the index of the $222L$ theory itself one can observe that it is equal to the index of the $usp(4)$ theory with four fundamental
hypermultiplets, one antisymmetric hypermultiplet and a free decoupled hypermultiplet.
The four fundamental hypers form a fundamental representation of $SO(8)$ flavor group,
which in terms of $SU(2)^4$ maximal sub-group is given by
\be
{\bf 8}={\bf 2}_e\times{\bf 2}_b+{\bf 2}_c\times{\bf 2}_d\,.
\ee Since the antisymmetric representation is real there is an $sp(1)$ symmetry rotating
the two corresponding half-hypermultiplets which we parametrize by fugacity $a$. Then by explicit computation the index
is given by (we bring the technical details in appendix~\ref{secapp})
\be\label{uspsu2}
{\cal I}_{222L}(c,d,e; a,b)=\frac{1}{1-\t a^{\pm1}}\,
{\cal I}_{usp(4)+4f+1a}(b,c,d,e;a)\,.
\ee  Here again we can define different $a$ fugacity for the free component and the
interacting one resurrecting our prescription. Let us observe that, similarly to what happened
in the previous subsection, $usp(4)$ theory has an ``accidental'' enhancement of flavor 
symmetry with the addition of $sp(1)$ and thus the $222L$ quiver captures this fact more naturally
than the $2222$ one.

 The pattern emerging from the two examples we discussed is that some of the bad theories 
which are believed to have good physical description have two useful properties. First, the naive flavor symmetry is enhanced:
in the case of ${\cal N}=4$ SYM that was the bigger R-symmetry of the SUSY algebra, and in the $usp(4)$ example of this section the 
$sp(1)$ symmetry rotating the antisymmetric hypermultiplet. Moreover, these bad theories appear as an interacting component of an ugly
 theory. Thus if such a situation occurs, the index of the bad theory can be directly inferred from the index of the ugly one. 
We will use this observation in the next section to compute an index of a bad theory which we did not know a-priori.
  
\
 
The HL index of the $usp(4)$ theory is actually equal~\cite{Gadde:2011uv} 
to the Hilbert series 
of the Higgs branch of the same theory~\cite{Benvenuti:2010pq}. As such, it counts the holomorphic functions on the two-instanton
moduli space of $SO(8)$ on ${\mathbb R}^4={\mathbb C}^2$.\footnote{In general it is believed
that the Hilbert series of the $usp(2k)$ theory with one antisymmetric hypermultiplet and 
four fundamental hypermultiplets is related to the moduli space of $k$ instantons.}
In this 
context the decoupled free hyper-multiplet in~\eqref{uspsu2}
is naturally interpreted as the center of mass degree of freedom~\cite{Aharony:2007dj,Benvenuti:2010pq}.
Moreover the extra $sp(1)$ symmetry 
is naturally interpreted as the one rotating the two complex planes of ${\mathbb C}^2$~\cite{Aharony:2007dj,Benvenuti:2010pq}. 
By the same logic HL index of $N_f=4$ $SU(2)$ theory 
describes the one-instanton moduli space. 
In one-instanton case the $sp(1)$ symmetry only affects the center of mass degrees of freedom
and thus the centered moduli space did not have this symmetry acting non-trivially on it:
 for multi-instantons
it plays the role for both center of mass and the centered moduli spaces.

It is interesting to note that the coefficient of
the leading second order singularity at $a=\t$ in~\eqref{uspsu2} is
\be
\frac{\left(\t +\t ^3\right)^2 \left(1+17 \t ^2+48 \t ^4+17 \t ^6+\t ^8\right)^2}{2 \left(-1+\t ^2\right)^{22}}=
\half \frac{\t^2}{(1-\t^2)^2}\,\left({\cal I}_{N_f=4,SU(2)}\right)^2\,,
\ee where for simplicity we set all the flavor fugaicities to one. Here ${\cal I}_{N_f=4,SU(2)}$ is the index of $N_f=4$ $SU(2)$ SYM.
The above result is thus physically naturally interpreted as ``going''
to the locus in the two-instanton moduli space where the two instantons are far away from each other and the
moduli space looks just as a product of moduli spaces of two one-instantons.
Moreover, to the lowest orders in $\t$ expansion the index of the $usp(4)$ theory looks like 
the symmetric product of indices of two $N_f=4$ $SU(2)$ SCFTs,
\be
&&\frac{1}{1-\t x^{\pm1}}\,
{\cal I}_{usp(4)+4f+1a}(\t)=\\
&&\qquad\half\left[\left({\cal I}_{N_f=4,SU(2)}(\t)\frac{1}{1-\t^{} x^{\pm1}}\right)^2+{\cal I}_{N_f=4,SU(2)}(\t^2)\frac{1}{1-\t^2 x^{\pm2}}\right]+O(\t^4)\,.\nonumber
\ee In particular up to order $\t^4$ the terms appearing in the index
describe generators of the Higgs branch. 
Thus, the parameters of the two-instanton moduli space are those of the 
symmetric product of two one-instantons and the details of the geometry, encoded
in the Higgs branch constraints, are different between the two.

\subsection{The index of rank two $E_6$ SCFT}

\begin{figure}
 \begin{center}
 \includegraphics[scale=0.35]{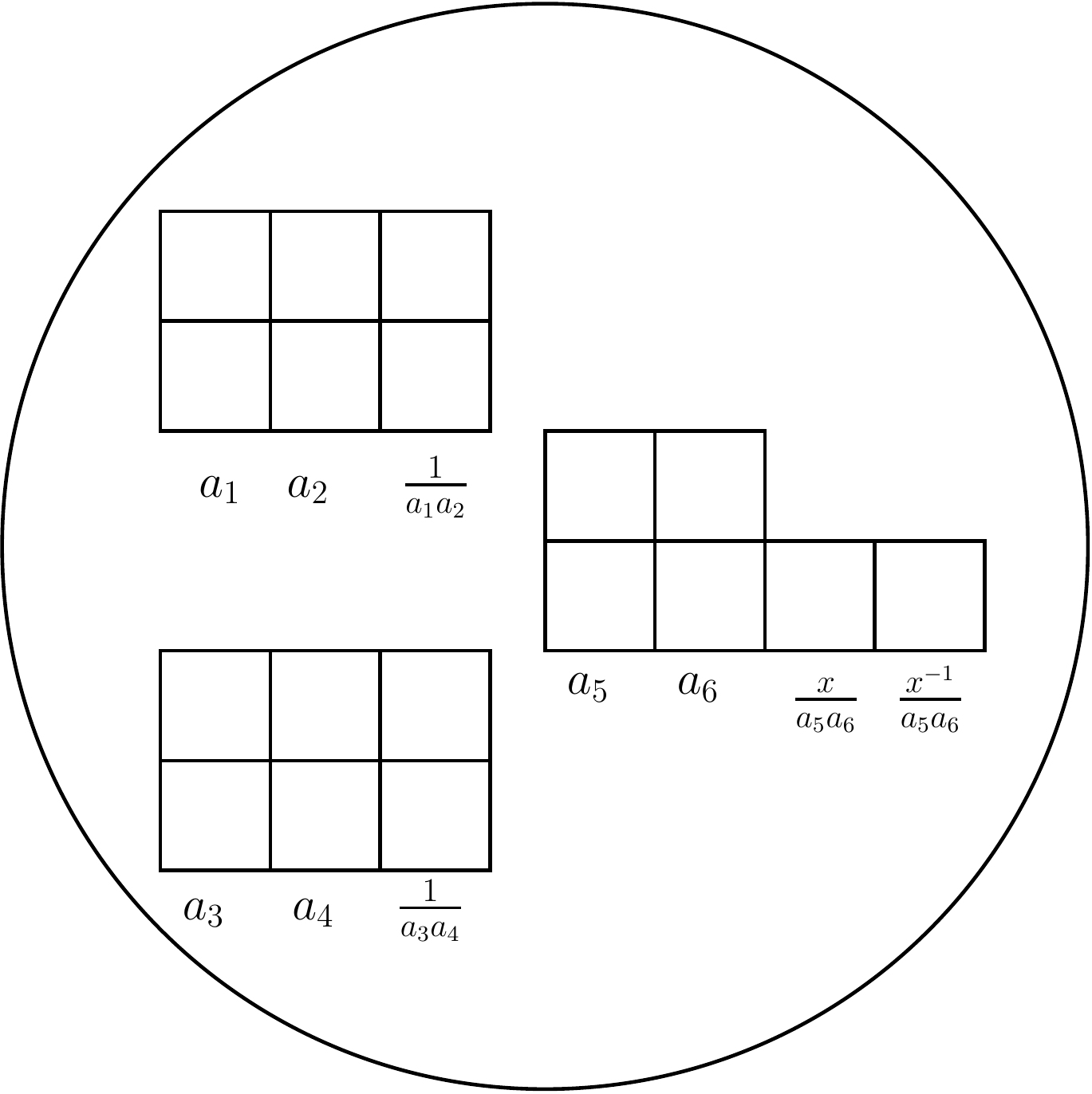}
 \end{center}
 \caption{A vertex relevant for the two-instanton moduli space of $E_6$.\label{2instfig}}
 \end{figure}

It is believed~\cite{Benini:2009gi,Moore:2011ee} that the $A_5$ theory corresponding to three punctured sphere with three
identical punctures with $SU(3)$ flavor symmetry (two rows with three boxes each) describes
an SCFT with $E_6$ flavor symmetry and rank two. This is a bad theory,
\be
\upsilon=(2,0,-2,2,0,-2)\,,
\ee with representation $\lambda=(\ell,\ell,\ell,0,0,0)$ causing the divergency. 
Thus, the prescription of~\cite{Gadde:2011uv}  here gives a divergent result. 
One can try to obtain the index of this theory
by the residue technology described in this paper but at the final step (the theory depicted in figure~\ref{2instfig})
a double pole is encountered. However, based on the results of the previous 
subsection, and  in analogy to ${\cal N}=4$ SYM and the rank two $SO(8)$ SCFT, it is tempting to suggest that the theory of figure~\ref{2instfig} is \textit{itself}
the rank two $E_6$ SCFT with decoupled hyper-multiplet.  
The theory is an ugly one,
\be
\upsilon=(2,0,-1,1,0,-2)\,,
\ee with representation $\lambda=(1,1,1,0,0,0)$ giving contribution at order $\t$.
The flavor symmetry here is $SU(3)\times SU(3)\times S(U(2)\times U(2))$ and we again expect 
the ugly theory to be equal to the bad one with an addition of a free hypermultiplet. The rank of this theory is
two as desired and the Coulomb branch generators have dimensions three and six.
Using the general definitions of~\cite{Gadde:2011uv} reviewed here in section~\ref{boxsec} the index of this theory is given by
\be\label{expe6a}
&&{\cal I}={\cal N}_6\, {\cal K}_1(a_1,a_2)\,{\cal K}_1(a_3,a_4)\,
{\cal K}_2(a_5,a_6,x)\,\sum_\lambda
\frac{
\psi_\lambda(\t a_5,\t^{-1} a_5,\t a_6,\t^{-1} a_6, \frac{x}{a_5a_6},\frac{x^{-1}}{a_5a_6}|\t)
}
{\psi_\lambda(\t^{-5},\t^{-3},\t^{-1},\t^{1},\t^{3},\t^{5}|\t)}\,\times\nonumber\\
&&\psi_\lambda(\t a_1,\t^{-1} a_1,\t a_2,\t^{-1} a_2,\t \frac{1}{a_1a_2},\t^{-1} \frac{1}{a_1a_2}|\t)
\psi_\lambda(\t a_3,\t^{-1} a_3,\t a_4,\t^{-1} a_4,\t \frac{1}{a_3a_4},\t^{-1} \frac{1}{a_3a_4}|\t)\,.
\nonumber\\
\ee
Here $\lambda=(\lambda_1,\cdots,\lambda_5,0)$ and ($b_3\equiv\frac{1}{b_1b_2}$)
\be
&&{\cal K}_1(b_1,b_2)=\prod_{\ell=1}^2\prod_{i,j=1}^3\frac{1}{1-\t^{2\ell} b_i/b_j},\nonumber\\
&&{\cal K}_2(b_1,b_2,x)=\left[\prod_{\ell=1}^2\prod_{i,j=1}^2\frac{1}{1-\t^{2\ell} b_i/b_j}\right]\times\\
&&\qquad\frac{1}{(1-\t^2)^2}\frac{1}{1-\t^2x^{\pm2}}
\frac{1}{1-\t^3(b_1^2b_2x^{\pm1})^{\pm1}}\frac{1}{1-\t^3(b_1b_2^2x^{\pm1})^{\pm1}}
\,.\nonumber
\ee
Expanding this index in power series in $\t$ to lowest orders we obtain 
\be\label{expe6}
{\cal I}=&&
\frac{1}{1-x^{\pm1}\t}\biggl[
1+\left[\chi^{E_6}_{78}(a)+\chi^{su(2)}_{3}(x)\right]\,\t^2+
\chi^{E_6}_{78}(a)\chi^{su(2)}_{2}(x)\,\t^3+\\
&&\qquad\qquad+
\left[\chi^{E_6}_{78}(a)\chi^{su(2)}_3(x)+\chi^{E_6}_{Sym^2 78}(a)+\chi_{Sym^2 {\bf 3}_{su(2)}}(x)-1\right]\,\t^4+\nonumber\\
&&\qquad\qquad+\left[\chi^{E_6}_{78}(a)\chi^{su(2)}_{2}(x)\,
\left[\chi^{E_6}_{78}(a)+\chi^{su(2)}_{3}(x)\right]-\chi^{E_6}_{27}(a)\,\chi^{E_6}_{\overline{27}}(a)\,\chi^{su(2)}_{2}(x)\right]\,\t^5
+\dots
\biggr]\,,\nonumber
\ee 
where we have
\be
&&\chi^{E_6}_{27}(a)=\chi^{su(3)}_{3}(a_1,a_2)\chi^{su(3)}_{\overline{3}}(a_3,a_4)+
\chi^{su(3)}_{3}(a_3,a_4)\chi^{su(3)}_{\overline{3}}(a_5,a_6)+
\chi^{su(3)}_{3}(a_5,a_6)\chi^{su(3)}_{\overline{3}}(a_1,a_2),\nonumber\\
&&\chi^{E_6}_{78}(a)=\chi^{su(3)}_{8}(a_1,a_2)+\chi^{su(3)}_{8}(a_3,a_4)+\chi^{su(3)}_{8}(a_5,a_6)+\\&&
\qquad\qquad\chi^{su(3)}_{3}(a_1,a_2)\chi^{su(3)}_{3}(a_3,a_4)\chi^{su(3)}_{3}(a_5,a_6)+
\chi^{su(3)}_{\bar 3}(a_1,a_2)\chi^{su(3)}_{\bar 3}(a_3,a_4)\chi^{su(3)}_{\bar3}(a_5,a_6)\,,\nonumber\\
&&\chi^{E_6}_{Sym^2 78}(a)=\half\left((\chi^{E_6}_{78}(a))^2+\chi^{E_6}_{78}(a^2)\right)\,.\nonumber
\ee
The flavor fugacities $a_i$  form characters of $E_6$ representations
providing a highly non-trivial check of our suggestion.
Stripping off the free hypermultiplet we conjecture that we get the index of the rank two $E_6$
SCFT. The generators of the Higgs branch are visible in the expression at orders $\t^{2,3}$, and at orders $\t^{4,5}$
we have  constraints appearing. As was the case in the previous example in addition to $E_6$ flavor symmetry we have more flavor symmetry: here it is $SU(2)$ parametrized by $x$.

\

As we mentioned several times before, it was argued in~\cite{Gadde:2011uv} that the HL index 
for quivers with topology of a sphere is equivalent to the Hilbert series of
the Higgs branch. The proof is explicit for theories with Lagrangian description.
The argument can be extended also to theories related to Lagrangian by S-dualities.
The model in this section can not be connected to Lagrangian theories by S-duality
(i.e. it can not be glued to another theory by gauging a maximal puncture due to the simple fact that it lacks a maximal puncture).
However, we can still conjecture that also here the HL index  is equivalent to the 
Hilbert series of the Higgs branch. One motivation for this statement is that we obtain the index by a chain
of residue computations each of which can be interpreted as gauging a flavor symmetry.
Thus, it would be extremely interesting to study in detail this index because of its possible interpretation
as the Hilbert series of the Higgs branch and subsequently as the Hilbert series of the two-instanton moduli space
of $E_6$. 

Let us comment that the above expression~\eqref{expe6}, up to few lowest  orders, can be understood 
as a symmetric product of two rank one $E_6$ SCFT indices.
The Higgs branch and its constraints for this theory were studied in~\cite{Gaiotto:2008nz} and an elegant 
expression for the Hilbert series was given recently in~\cite{Benvenuti:2010pq}.\footnote{
Such expressions are known in mathematical literature, e.g.~\cite{Garfinkle}.
We thank Yuji Tachikawa for pointing this out to us.}
 This information is also neatly encoded in the index which  was computed in~\cite{Gadde:2010te} and 
is given by
\be
{\cal I}^{rank\,1}_{E_6}(a,\t)=1+\chi^{E_6}_{78}(a)\t^2+(\chi^{E_6}_{Sym^2 78}(a)-\chi^{E_6}_{27}(a)\,\chi^{E_6}_{\overline{27}}(a)+\chi^{E_6}_{78}(a))\t^4+\dots\,.
\ee Taking the symmetric product of this tensored with a free hypermultiplet we obtain
\be
\frac{1}{1-\t x^{\pm1}}{\cal I}^{rank\,2}_{E_6}(a,\t)=\half\left[\left({\cal I}^{rank\,1}_{E_6}(a,\t)\frac{1}{1-\t x^{\pm1}}\right)^2+{\cal I}^{rank\,1}_{E_6}(a^2,\t^2)
\frac{1}{1-\t^2 x^{\pm2}}\right]+O(\t^4)\,.\nonumber\\
\ee  Also here the generators of the two-instanton case seem to be inherited from the symmetric product
of two one-nstantons.
At higher orders in $\t$ the expressions start to deviate from each other
due to the different constraint systems.

\

\begin{figure}
\begin{center}
\includegraphics[scale=0.35]{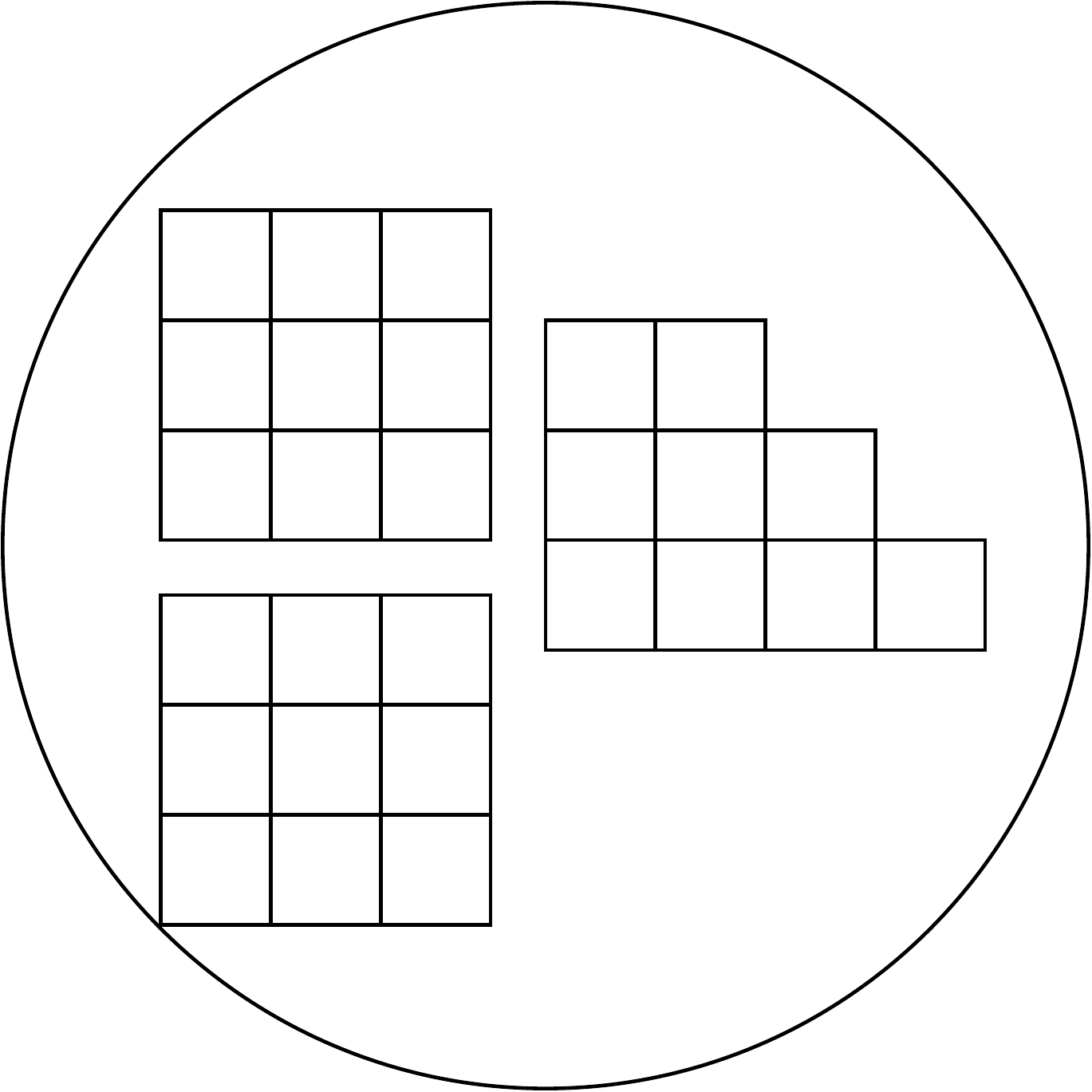}
\end{center}
\caption{A vertex relevant for the three-instanton moduli space of $E_6$.\label{3instfig}}
\end{figure}

Of course this method can be applied to obtain higher rank $E_6$ theories
and also higher rank $E_{7,8}$ models (rank one theories correspond to Minahan-Nemeschansky SCFTs~\cite{Minahan:1996fg,Minahan:1996cj} 
index of which is computed in~\cite{Gadde:2011ik,Gadde:2011uv}).
The prescription for $E_6$ would be to consider an $A_{3k-1}$ theory with two rectangular $SU(3)$ punctures
and third puncture obtained from $SU(3)$ rectangular one by lowering a box from the top right corner.
The rank three case is illustrated in figure~\ref{3instfig}. These theories have rank $k$ and we 
suggest that they are the same as the $A_{3k-1}$ theories with three rectangular $SU(3)$ punctures
with an addition of free hypermultiplet. We bring in the appendix a first order check of this proposal
for the rank three case of $E_6$ and the rank two case of $E_7$. It is also interesting to inquire whether our rank two $E_6$ index~\eqref{expe6a} has a nice manifestly
$E_6$ 
covariant closed form on par with the expression for the rank one Hilbert series appearing in~\cite{Benvenuti:2010pq,Keller:2011ek}.

\

\section*{Acknowledgments}

We would like to thank Guido Festuccia, Abhijit Gadde,  Leonardo Rastelli, and
Yuji Tachikawa for very useful comments and discussions.
SSR would like to thank the  HET group at the Weizmann Institute and KITP for hospitality during different stages of this project.
The research of SSR was supported in part by NSF grant PHY-0969448. 
The work of DG is supported in part by 
NSF grant NSF PHY-0969448 and in part
by the Roger Dashen membership in 
the Institute for Advanced Study.
This research was also supported in part by the NSF grant PHY-1125915.

\

\appendix

\section{Technical details}\label{secapp}

\subsection*{Examples of Hilbert series}

Let us give several simple examples of Hilbert series, i.e counting holomorphic functions
on complex manifolds. The Hilbert series counts holomorphic functions
giving the same weight to functions of same ``degree''. Simplest example is 
functions on the complex plane ${\mathbb C}$. Such functions are generated by
polynomials and the Hilbert series is
\be
{\cal H}=1+t+t^2+t^3+\cdots=\frac1{1-t}\,.
\ee Here monomials of degree $k$, $z^k$, receive weight $t^k$. For moduli space 
of instantons in ${\mathbb R}^4\sim {\mathbb C}^2$ the relevant series is 
\be
{\cal H}=\frac1{1-x^{\pm1}t}\,,
\ee where we further refined the series by parameter $x$ giving $z_1^k$ weight $x t^k$ 
and $z_2^k$ weight $x^{-1}t^k$. The parameter $x$ can be viewed as $SU(2)$ fugacity
rotating the two copies of the complex plane. This is also of course the HL index of a
free hypermultiplet and the Higgs branch Hilbert series. The only field in the hypermultiplet
contributing to both objects is a scalar $q$. The fugacity $x$ couples to  the $sp(1)$ symmetry rotating the 
half-hypers. 

Our second example is Hilbert series of two complex planes 
glued at the origin, ${\cal A}=\{(z_1,\,z_2)\in{\mathbb C}\times {\mathbb C}|\,z_1\,z_2=0\}$.\footnote{ We are grateful to N.~Seiberg for pointing out  inconsistencies in this
subsection in the previous version of the paper.}
The Hilbert series is thus given by,
\be
{\cal H}=\frac{1}{1-t_1}+\frac{1}{1-t_2}-1=\frac{1-t_1t_2}{(1-t_1)(1-t_2)}\,,
\ee Here $t_i$ couple to the coordinates $z_i$ of the two copies of the complex plane.
The constraint $1-t_1t_2$ implements the fact that the two planes are glued at the origin, $z_1\,z_2=0$.\footnote{Topologically equivalent way to view  ${\cal A}$ is  as
${\mathbb C}/\{0\}$, {\it i.e.} an annulus. In the language of ${\mathbb C}/\{0\}$ the interpretation is as follows.
Holomorphic functions on this space are generated by integer, positive
and negative, powers of $z$. Here $t_1$ couples to $z$ and $t_2$ to $1/z$.
 The constraint 
$1-t_1t_2$ implements the fact that $(z)(1/z)=1$.}
 Taking $t_i\to x^{\pm1}t$
we can write the Hilbert series as,
\be
{\cal H}=\frac{1-t^2}{1-x^{\pm1}t}\,.
\ee Here again $x$ can be thought of as an $SU(2)$ fugacity.
 The 
denominator comes from the hypermultiplet and the numerator from the vector multiplet
in the HL index language.  We have a hypermultiplet with a quadratic constraint.

Note that in both cases, ${\mathbb C}$ and
${\cal A}$, the degree of the singularity at $t=1$ of the un-refined series, $x=1$
is equal to one. This degree is interpreted as the complex dimension of the underlying space 
which in both cases is equal to one. The difference in the topology is encoded in the different
numerators in the two cases. Physically the different
topology comes from the different constraint system, or
in the HL index language simply because of the different matter content.

\

\subsection*{The index of a usp(4) theory and the $222L$ theory}

Let us compute the HL index of the $usp(4)$ gauge theory we encountered in the bulk 
of the paper. This theory has four hypermultiplet in the fundamental representation (${\bf 4}$)
of the gauge group and a single hypermultiplet in the antisymmetric representation (${\bf 5}$).
The adjoint representation of $usp(4)$ has dimension ten.
The characters of these representations are
\be
&&\chi_{\bf 4}=x_1+x_1^{-1}+x_2+x_2^{-1}\,,\qquad
\chi_{\bf 5}=1+x_1x_2+\frac1{x_1x_2}+\frac{x_1}{x_2}+\frac{x_2}{x_1}\,,\\
&&\chi_{{\bf {10}}}=2+x_1x_2+\frac1{x_1x_2}+\frac{x_1}{x_2}+\frac{x_2}{x_1}
+x_1^2+x_2^2+x_1^{-2}+x_2^{-2}\,.\nonumber
\ee The theory has $SO(8)$ flavor symmetry rotating the fundamental half-hypers and
an $sp(1)$ flavor symmetry rotating the half-hypers of the antisymmetric hypermultiplet. 
The HL index of this theory is then given by the following matrix integral
\be
&&{\cal I}_{usp(4)+4f+1a}(b,c,d,e;a)=\\
&&\qquad \oint \frac{dx_1}{2\pi i x_1}\oint \frac{dx_2}{2\pi i x_2}\Delta_{usp(4)}({\bf x})
PE\left[\t\chi^{SO(8)}_8\,
\chi_{\bf 4}({\bf x})+\t(a+a^{-1})\chi_{\bf 5}({\bf x})\right]\,
PE\left[-\t^2\chi_{{\bf{ 10}}}({\bf x})\right]\,,\nonumber
\ee where
\be
&&\chi^{SO(8)}_8=(e+e^{-1})(b+b^{-1})+(c+c^{-1})(d+d^{-1})\,,\\
&&\Delta_{usp(4)}({\bf x})=\frac1{8x_1^2x_2^2} \, (x_1-x_2)^2(x_1-x_1^{-1})^2(x_2-x_2^{-1})^2
(1-x_1x_2)^2\,,
\nonumber
\ee with the latter being the Haar measure of $usp(4)$. Here $PE[\cdot]$ is the plethystic exponent,
\be
PE[f(x,y,\cdots)]=\exp\left[\sum_{\ell=1}^\infty\frac1{\ell}\,f(x^\ell,y^\ell,\cdots)\right]\,.
\ee On the other hand, following the prescription reviewed in section~\ref{boxsec} the index of the $222L$ theory 
is given by,
\be
&&{\cal I}_{222L}(c,d,e; a,b)={\cal N}_4\, {\cal K}_1(c)\,{\cal K}_1(d)\,{\cal K}_1(e)\,{\cal K}_2(a,b)\,
\sum_\lambda
\frac{\psi_\lambda(\t b,\t^{-1} b,b^{-1}a,b^{-1}a^{-1}|\t)
}
{\psi_\lambda(\t^{-3},\t^{-1},\t^{},\t^{3}|\t)}\,\times\\
&&\qquad
\psi_\lambda(\t c,\t^{-1} c,\t c^{-1},\t^{-1} c^{-1}|\t)
\psi_\lambda(\t d,\t^{-1} d,\t d^{-1},\t^{-1} d^{-1}|\t)
\psi_\lambda(\t e,\t^{-1} e,\t e^{-1},\t^{-1} e^{-1}|\t)\,.\nonumber
\ee Here we have defined ($b_1=b,\;b_2=1/b$)
\be
&&{\cal K}_1(b)=\prod_{\ell=1}^2\prod_{i,j=1}^2\frac{1}{1-\t^{2\ell} b_i/b_j}\,,\\
&&{\cal K}_2(a,b)=\frac{1}{(1-\t^2)^3(1-\t^4)(1-\t^3 b^{\pm2}a^{\pm1})(1-\t^2 a^{\pm2})}\,.\nonumber
\ee  Comparing the above expressions order by order in $\t$ we conclude that
\be
{\cal I}_{222L}(c,d,e; a,b)=\frac{1}{1-\t a^{\pm1}}\,
{\cal I}_{usp(4)+4f+1a}(b,c,d,e;a)\,.
\ee Evaluating this index gives,
\be
{\cal I}_{222L}(1,1,1; 1,1)=\frac{1}{(1-\t)^2}\left[1+31 \t ^2+56 \t ^3+495 \t ^4+1468 \t ^5+6269 \t ^6+\cdots\right]\,.
\ee Here at $\t^2$ order  $31=28+3$ with $28$ being the adjoint of $SO(8)$ and $3$ adjoint
of $SU(2)$. At $\t^3$ order   $56=2\times 28$ with $2$ being fundamental of $SU(2)$ and $28$ 
adjoint of $SO(8)$: and so on.

\

\subsection*{The index of rank three $E_6$ SCFT}
Let us give the expression for the index of rank two $E_6$ theory. 
The Riemann surface is depicted in figure~\ref{3instfig}.
 The association of flavor fugacities to columns
of the auxiliary Young diagrams is as in figure~\ref{2instfig}
with the only difference being the box we lower: its fugacity is $\frac{x^{-2}}{a_5a_6}$.
Here the index is given by
\be
&&{\cal I}={\cal N}_9\, {\cal K}_1(a_1,a_2)\,{\cal K}_1(a_3,a_4)\,
{\cal K}_2(a_5,a_6,x)\,\times\\
&&\qquad\sum_\lambda
\frac{
\psi_\lambda(\t^2 a_5,\t^{-2} a_5,a_5,\t^2 a_6,\t^{-2} a_6,a_6,
 \t \frac{x}{a_5a_6}, \t^{-1}\frac{x}{a_5a_6},\frac{x^{-2}}{a_5a_6}|\t)
}
{\psi_\lambda(\t^{-8},\t^{-6},\t^{-4},\t^{-2},1,\t^2,\t^{4},\t^{6},\t^{8}|\t)}\,\times\nonumber\\
&&\qquad
\psi_\lambda(\t^2 a_1,\t^{-2} a_1,a_1,\t^2 a_2,\t^{-2} a_2,a_2,\t^2 \frac{1}{a_1a_2},\t^{-2} \frac{1}{a_1a_2},\frac{1}{a_1a_2}|\t)
\nonumber\\
&&\qquad
\psi_\lambda(\t^2 a_3,\t^{-2} a_3,a_3,\t^2 a_4,\t^{-2} a_4,a_4,\t^2 \frac{1}{a_3a_4},\t^{-2} \frac{1}{a_3a_4}, \frac{1}{a_3a_4}|\t)\,.
\nonumber
\ee
Here $\lambda=(\lambda_1,\cdots,\lambda_8,0)$ and ($b_3\equiv\frac{1}{b_1b_2}$)
\be
&&{\cal K}_1(b_1,b_2)=\prod_{\ell=1}^3\prod_{i,j=1}^3\frac{1}{1-\t^{2\ell} b_i/b_j}\,,\\
&&{\cal K}_2(b_1,b_2,x)=\left[\prod_{\ell=1}^3\prod_{i,j=1}^2\frac{1}{1-\t^{2\ell} b_i/b_j}\right]
\frac{1}{(1-\t^2)}
\frac{1}{1-\t^3(b_1^2b_2x^{-1})^{\pm1}}\frac{1}{1-\t^3(b_1b_2^2x^{-1})^{\pm1}}\times\nonumber\\
&&\qquad
\frac{1}{(1-\t^2)(1-\t^4)}
\frac{1}{1-\t^5(b_1^2b_2x^{-1})^{\pm1}}\frac{1}{1-\t^5(b_1b_2^2x^{-1})^{\pm1}}\times\nonumber\\
&&\qquad\frac{1}{1-\t^4(b_1^2b_2x^{2})^{\pm1}}\frac{1}{1-\t^4(b_1b_2^2x^{2})^{\pm1}}
\frac{1}{1-\t^3x^{\pm3}}
\,.\nonumber
\ee
Expanding this index in power series in $\t$ to lowest orders we obtain 
\be
{\cal I}=&&
\frac{1}{1-x^{\pm1}\t}\biggl[
1+(\chi^{E_6}_{78}(a)+\chi^{su(2)}_{3}(x))\,\t^2+\dots
\biggr]\,.\nonumber
\ee Consistently with our claim this index organizes itself in $E_6$ representations. \footnote{
The higher order corrections here are technically hard to compute since the complexity of the 
HL polynomials grows exponentially with the rank of the group. We used the support for 
Hall-Littlewood polynomials in {\it{Sage}} ({\it{http://www.sagemath.org}}) to generate the polynomials.}
To obtain the index of the rank three $E_6$ SCFT one strips off the free hypermultiplet.

\

\subsection*{The index of rank two $E_7$ SCFT}

\begin{figure}
\begin{center}
\includegraphics[scale=0.35]{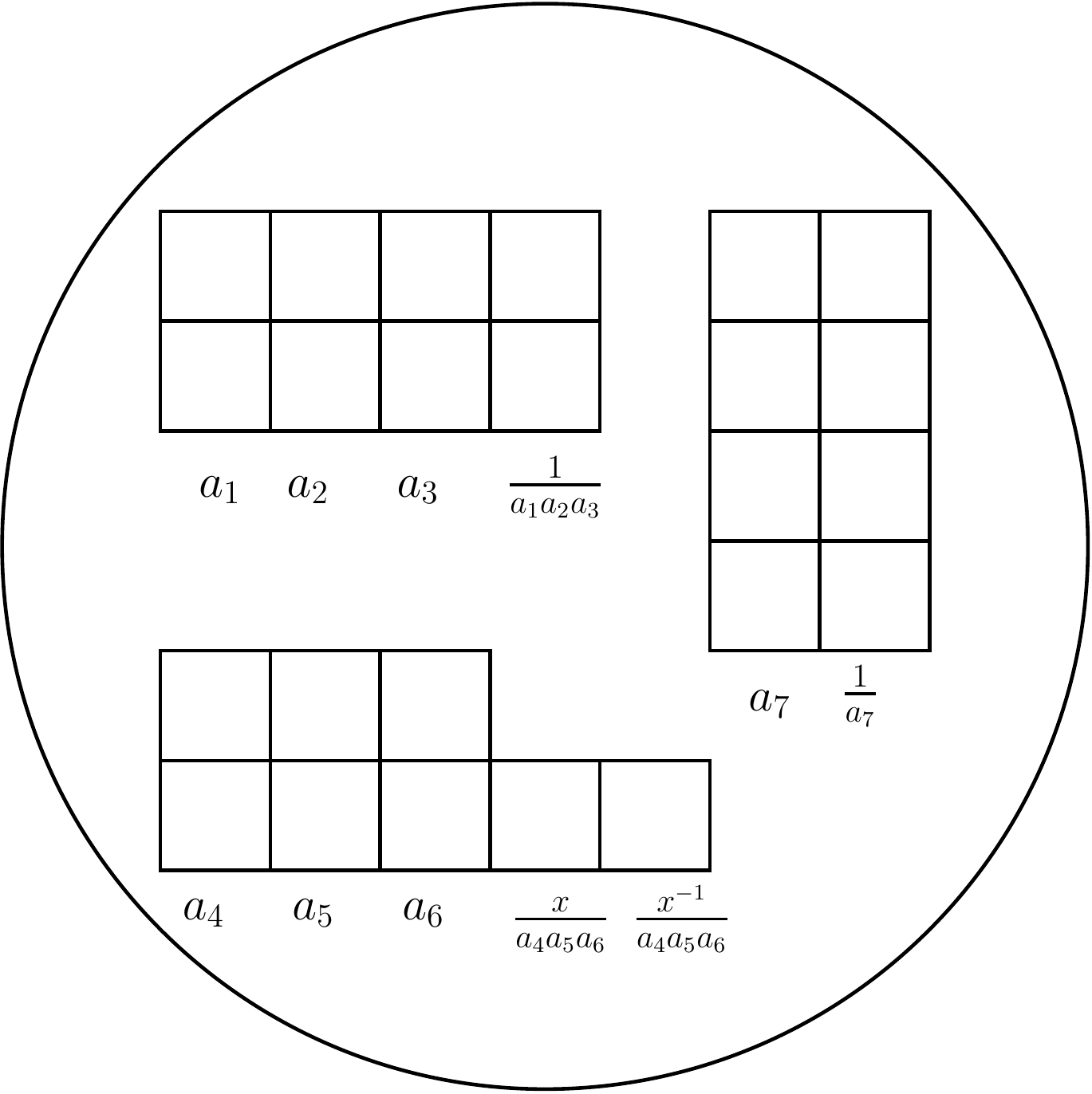}
\end{center}
\caption{The sphere with three punctures corresponding to the rank two SCFT with $E_7$ 
flavor symmetry and decoupled hypermultiplet.\label{E7fig}}
\end{figure}

Let us consider the rank two SCFT with $E_7$ flavor symmetry.
The relevant $A_7$ rank two quiver theory is depicted on figure~\ref{E7fig}.
Here the index is given by
\be
&&{\cal I}={\cal N}_8\, {\cal K}_1(a_1,a_2,a_3)\,{\cal K}_2(a_4,a_5,a_6,x)\,
{\cal K}_3(a_7)\,\times\\
&&\qquad\sum_\lambda
\frac{
\psi_\lambda(\t^3 a_7,\t^{-3} a_7,\t a_7,\t^{-1} a_7,\t^3 a_7^{-1},\t^{-3} a_7^{-1},\t a_7^{-1},\t^{-1} a_7^{-1}|\t)
}
{\psi_\lambda(\t^{-7},\t^{-5},\t^{-3},\t^{-1},\t,\t^{3},\t^{5},\t^{7}|\t)}\,\times\nonumber\\
&&\qquad
\psi_\lambda(\t a_1,\t^{-1} a_1,\t a_2,\t^{-1} a_2,\t^{} a_3,\t^{-1} a_3,\t \frac{1}{a_1a_2a_3},\t^{-1} \frac{1}{a_1a_2a_3}|\t)
\nonumber\\
&&\qquad
\psi_\lambda(\t a_4,\t^{-1} a_4,\t a_5,\t^{-1} a_5,\t^{} a_6,\t^{-1} a_6,\frac{x}{a_4a_5a_6},\frac{x^{-1}}{a_4a_5a_6}|\t)\,.
\nonumber
\ee
Here $\lambda=(\lambda_1,\cdots,\lambda_7,0)$ and ($b_4\equiv\frac{1}{b_1b_2b_3}$)
\be
&&{\cal K}_3(b)=\frac{1}{(1-\t^2)^2(1-\t^4)^2(1-\t^6)^2(1-\t^8)^2}\prod_{\ell=1}^4\frac{1}{1-\t^{2\ell} b^{\pm2}}\,,\\
&&{\cal K}_2(b_1,b_2,b_3,x)=\frac{1}{(1-\t^2)^2(1-\t^4)^3}\left[\prod_{\ell=1}^2\prod_{i,j=1}^3\frac{1}{1-\t^{2\ell} b_i/b_j}\right]
\prod_{i=1}^3\frac{1}{1-\t^3 (x^{-1}\,b_{i}/b_4)^{\pm1}}\times\nonumber\\
&&\qquad \prod_{i=1}^3\frac{1}{1-\t^3 (x^{}\,b_{i}/b_4)^{\pm1}}
\frac{1}{1-\t^2\,x^{\pm2}}\,,
\nonumber\\
&&{\cal K}_1(b_1,b_2,b_3)=\left[\prod_{\ell=1}^2\prod_{i,j=1}^3\frac{1}{1-\t^{2\ell} b_i/b_j}\right]\,.\nonumber
\ee
The expansion of the index is given by
\be
{\cal I}=\frac{1}{1-x^{\pm1}\t}\;\left[1+\left(\chi^{E_7}_{133}(a)+\chi^{su(2)}_3(x)\right)\t^2+\cdots\right]\,.
\ee Here the embedding of the flavor fugacities inside $E_7$ is
\be
{\bf 133}_{E_7}={\bf 3}_{su(2)}+{\bf 2}_{su(2)}\left({\bf 4}_1{\bf 4}_2+{\bf\overline{4}}_1
{\bf\overline{4}}_2\right)+{\bf 15}_1+{\bf 15}_2+{\bf 6}_1\,{\bf 6}_2\,,
\ee where indices $1$ and $2$ refer to the two $SU(4)$ groups parametrized by $(a_1,\,a_2,\,a_3)$
and by $(a_4,\,a_5,\,a_6)$. The $SU(2)$ is parametrized by $a_7$.

\begin{figure}
\begin{center}
\includegraphics[scale=0.35]{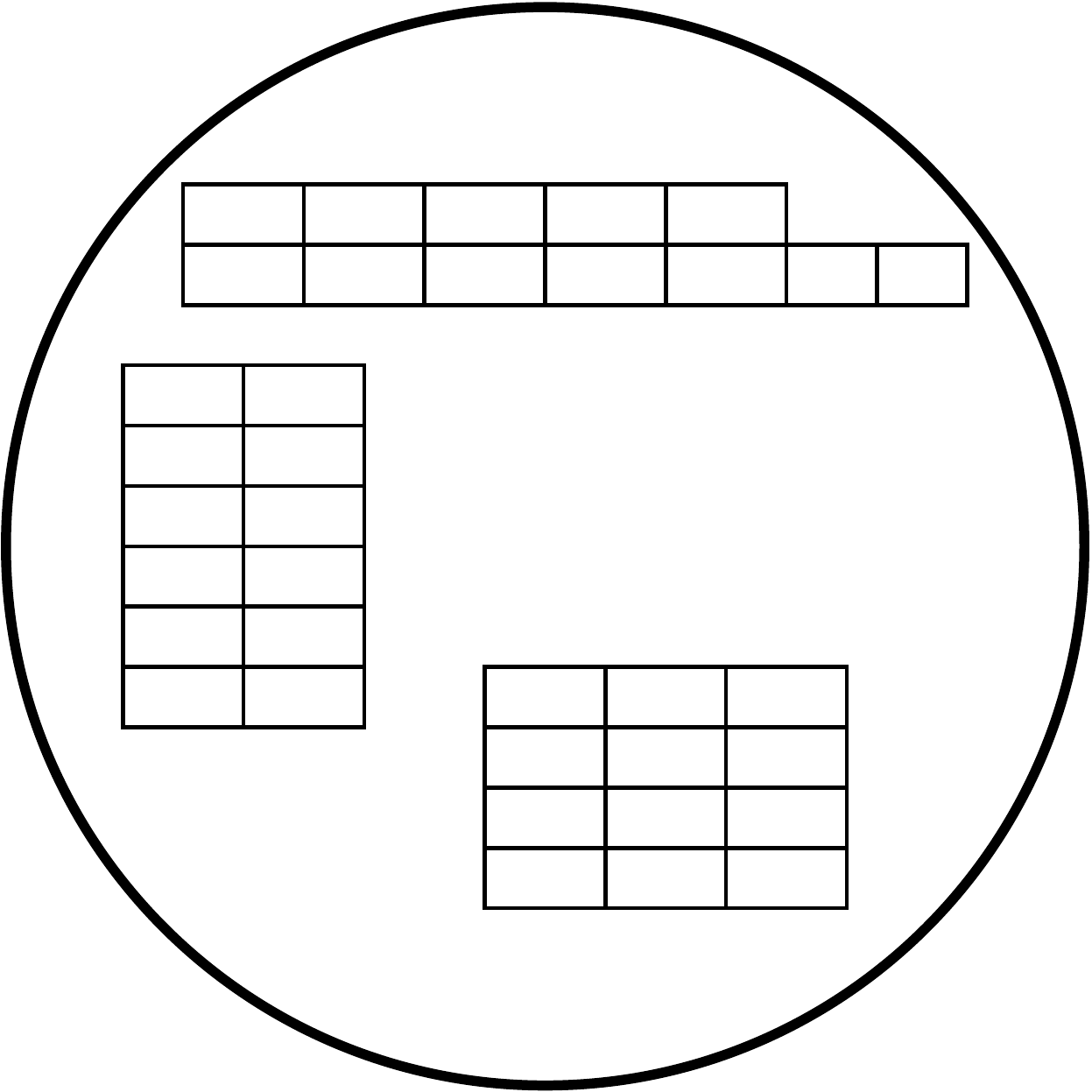}
\end{center}
\caption{The sphere with three punctures corresponding to the rank two SCFT with $E_8$ 
flavor symmetry and decoupled hypermultiplet.\label{E82fig}}
\end{figure}

Similar expression can be written also for the $E_8$ higher rank theories. For instance the rank two 
case is depicted in figure~\ref{E82fig}. The index can be calculated following our usual prescription.

\

\


\bibliography{sdualityMAC}

\bibliographystyle{JHEP}

\end{document}